\documentclass[preprint,amsmath,amssymb,nofootinbib]{revtex4}
\usepackage{graphicx}
\usepackage{slashed}
\usepackage{subfig}
\usepackage{float}
\usepackage[utf8]{inputenc}
\usepackage{color}
\usepackage{array}
\definecolor{red}{rgb}{1,0,0}

\newcommand{\tvect}[2]{%
  \ensuremath{\Bigl(\negthinspace\begin{smallmatrix}#1\\#2\end{smallmatrix}\Bigr)}}

\begin{document}

\title{Not-that-heavy Majorana neutrino signals at the LHC.}

\author{Luc\'{\i}a Duarte}
\email{lduarte@fing.edu.uy}
 \affiliation{Instituto de F\'{\i}sica, Facultad de Ingenier\'{\i}a,
 Universidad de la Rep\'ublica \\ Julio Herrera y Reissig 565,(11300) 
Montevideo, Uruguay.}

\author{Javier Peressutti}
\author{Oscar A. Sampayo}
\email{sampayo@mdp.edu.ar}

 \affiliation{Instituto de Investigaciones F\'{\i}sicas de Mar del Plata (IFIMAR)\\ CONICET, UNMDP\\ Departamento de F\'{\i}sica,
Universidad Nacional de Mar del Plata \\
Funes 3350, (7600) Mar del Plata, Argentina}

\begin{abstract}
We revisit the possibility of Majorana neutrinos production at the Large Hadron Collider (LHC) by studying the $pp \rightarrow l_i^{+} l_j^{+} +2~jets$ ($l_j\equiv e ,\mu$) process which, due to leptonic number violation, is a clear signature for intermediate Majorana neutrino contributions. The interactions between Majorana neutrinos and the Standard Model particles are obtained from an effective Lagrangian approach. Majorana neutrinos with masses of a few $GeV$ are long-lived neutral particles, and  we take advantage of its measurable decay length: in the same-sign dilepton channel, we  exploit this fact imposing cuts that reject the SM background, and analyze the distribution corresponding to the angle between the final leptons, using a forward-backward like asymmetry to study the effects of the different gauge invariant operators. 
We also study the $pp\rightarrow l_i^+ \nu \gamma$ process, which is dominant for low $m_N$ masses if tensorial one-loop generated new physics leading to a magnetic moment for the heavy neutrinos is present. This channel provides a powerful signal that could be observed at the LHC with the aid of non-pointing photons observables and cuts on the displacement between the prompt lepton and the photon in the final state. 
\end{abstract}

\maketitle

\section{Introduction}\label{sect:introduction}

While the discovery of neutrino oscillations and its interpretation in terms of non-zero neutrino masses remains as the most compelling evidence for the existence of physics beyond the Standard Model, many proposals have been made to explain the tiny neutrino masses, and the seesaw mechanism stays as one of the most straightforward means for solving this problem \cite{Minkowski:1977sc, Mohapatra:1979ia, Yanagida:1980xy, GellMann:1980vs, Schechter:1980gr, Kayser:1989iu}. The mechanism introduces right-handed sterile neutrinos that, as they do not have distinct particle and antiparticle degrees of freedom, can have a Majorana mass term leading to the known masses for the standard neutrinos, as long as the Yukawa couplings between the right handed Majorana neutrinos and the standard ones remain small. The Majorana mass term implies the possibility for lepton number violating (LNV, $\Delta L=2$) processes involving charged leptons.

The inquiry into the Majorana nature of neutrinos has led to dedicated searches for evidence of LNV at hadron colliders in the very well known same-sign dilepton (ss-dilepton) channel $pp \rightarrow l_i^{+} l_j^{+} +  2 ~jets$ (see \cite{Atre:2009rg, delAguila:2007qnc, delAguila:2008ir, Kovalenko:2009td, Alva:2014gxa, Das:2015toa, Dib:2015oka, Izaguirre:2015pga, Chakdar:2016adj,Dube:2017jgo} and references therein) recently including new production mechanisms \cite{Das:2016hof, Degrande:2016aje, Dev:2013wba}, and the chances to discover heavy Majorana neutrinos in $e^{+}e^{-}$, $e^{-}P$, $e^{-}\gamma$ and $\gamma \gamma$ colliders have been studied \cite{Ma:1989jpa, Buchmuller:1991tu, Hofer:1996cs, Peressutti:2011kx, Blaksley:2011ey, Duarte:2014zea, Antusch:2015mia, Bray:2005wv, Peressutti:2001ms, Peressutti:2002nf, Antusch:2016ejd}. Searches for heavy Majorana neutrinos in the ss-dilepton channel are currently being performed at the LHC \cite{Khachatryan:2016olu, Khachatryan:2015gha, Aad:2015xaa, Aad:2011vj, ATLAS:2012ak}.

In the naïve Type-I seesaw scenarios often studied, Yukawa couplings of order $Y \sim 1$ require a Majorana neutrino mass scale of order $M_{N} \sim 10^{15} ~GeV$ to account for a light $\nu$ mass compatible with the current oscillation data ($m_{\nu}\sim 0.01 eV$), and this fact leads to the decoupling of the Majorana neutrinos. On the other hand, for smaller Yukawa couplings $Y \sim (10^{-8}-10^{-6})$ sterile neutrinos with masses around $(1-1000) ~GeV$ could exist, but in the simplest Type-I seesaw scenario this leads to a negligible left-right neutrino mixing $U_{lN}^2 \sim m_{\nu}/M_N \sim (10^{-14}-10^{-10})$. Thus, as suggested in \cite{delAguila:2008ir}, the detection of Majorana neutrinos ($N$) via $\Delta L=2$ processes would be a signal of physics beyond the minimal seesaw mechanism. The Majorana neutrino interactions could be best described in a model independent approach based on an effective theory, considering Majorana neutrinos $N$ with negligible mixing with the $\nu_{L}$.

The effective interactions for the heavy Majorana neutrinos were early studied in \cite{delAguila:2008ir}, where the possible phenomenology of dimension 6 effective operators was introduced. The dimension 5 operators extending the low-scale Type-I seesaw were investigated in \cite{Aparici:2009fh}. Their phenomenology is addressed in recent works as  \cite{Ballett:2016opr} for the  Fermilab SBN program, and in \cite{Caputo:2017pit} concerning $N-$ Higgs interactions in hadron colliders. The dimension 7 effective $N$ operators are studied in \cite{Bhattacharya:2015vja, Liao:2016qyd}. The feasibility of observing heavy Majorana neutrinos with mass $m_{N}\lesssim 30 ~GeV$ in the future Belle-II and ILC experiments is studied in \cite{Yue:2017mmi}.  

The effective model we consider in this paper \cite{delAguila:2008ir} has been tested in the LHC Run I \cite{Aad:2011vj, ATLAS:2012ak} for Majorana neutrino masses above $100~GeV$, in events with high transverse momentum objects including two reconstructed leptons and jets, for $\sqrt{s}=7~TeV$. The data are found to be in agreement with the expected SM background, leading to limits in the effective couplings and new physics scale for some selected operators. If this kind of sterile neutrino exists for $m_N < m_W$, the produced jets in the final state $l_i^{+} l_j^{+} +  2 ~jets$ may not pass the cuts required to reduce backgrounds, as pointed out, for example, in ref. \cite{Dib:2016wge}.  

Indeed, for masses $m_N$ around a few $GeV$ the Majorana neutrino we are considering behaves as a long-lived neutral particle, with a measurable decay length. This gives us a new means for probing the effective new physics at this lower mass scale, taking advantage of displaced vertices techniques. Recent works use displaced vertices for studying heavy sterile neutrinos in the LHC \cite{Batell:2016zod, Izaguirre:2015pga, Helo:2013esa, Gago:2015vma, Cerdeno:2013oya, Caputo:2017pit, Antusch:2017hhu} and future colliders \cite{Blondel:2014bra, Antusch:2016vyf}.

In our previous work \cite{Duarte:2015iba} we found that for $m_{N}\lesssim 30 ~GeV$, the dominant neutrino plus photon $N\rightarrow \nu \gamma$ decay channel is given by the contribution of effective tensorial operators generated at one-loop level in the unknown underlying ultraviolet theory. As this channel cannot shed light on the Majorana or Dirac nature of heavy neutrinos, we first tackle the LNV same-sign dilepton signals for low Majorana neutrino masses neglecting the one-loop operators contribution. This enables us to test the capability to discern between the different gauge invariant operators contribution to the $pp \rightarrow l_i^{+} l_j^{+} +  2 ~jets$ process, using a forward-backward like asymmetry, and imposing displaced-vertices cuts that reject the SM background, in a scenario with no one-loop-generated operators contribution. 

In spite of not being a LNV signal, a study of the dominant neutrino plus photon decay channel is included in this paper, as it also was found that long-lived neutral radiatively decaying particles like the Majorana neutrino $N$, could explain the MiniBooNE \cite{AguilarArevalo:2007it, AguilarArevalo:2008rc} and SHALON \cite{Sinitsyna:2013hmn} anomalies \cite{Duarte:2015iba}, following sterile neutrino explanations for these experimental puzzles \cite{Gninenko:2009ks}. The radiative $N\rightarrow \nu \gamma$ channel can be observed by the signature of an isolated electromagnetic cluster together with missing transverse energy, where the photon originates in a displaced vertex. New physics searches involving displaced photons and missing transverse energy have been performed at the LHC \cite{Aad:2014gfa, Aad:2013oua}, mainly dedicated to SUSY searches, and searches with the same final state: lepton, photon and missing $E_{T}$ \cite{CMS:2015loa}.

The paper is organized as follows. In sections \ref{sect:eff_mod} and \ref{sect:bounds} we review the effective Lagrangian approach and the experimental bounds on the effective couplings. In Sec.\ref{sect:low_mN} we discuss the distinctive features of the low $m_N$ region for our signals. The same-sign dilepton plus jets process is studied in Sec.\ref{sect:ss_dilepton}, introducing the displaced leptons distance $L^{l^{+}l^{+}}$ and our results for the leptonic forward-backward like asymmetry $A^{l^{+}l^{+}}_{FB}$. In Sec.\ref{sect:photon_neutrino} we study the prompt lepton and neutrino plus photon process, presenting the non-pointing photon $z_{DCA}$ and displaced lepton-photon distance $L^{l^{+}\gamma}$ signal distributions. Our final remarks are given in Sec.\ref{sect:final}.

\subsection{Effective model}\label{sect:eff_mod}

While most of the past work has been focused on the investigation of heavy Majorana neutrinos that mix with the SM light neutrinos in the framework of low scale Type-I seesaw scenarios (see \cite{Atre:2009rg, delAguila:2007qnc}), the aim of our approach is to investigate the possible contributions of a heavy Majorana neutrino with negligible mixing to the SM $\nu_{L}$. Although the minimal framework capable of accommodating the measured light neutrino masses requires the introduction of two singlet Majorana fermions, in this work we will consider the most simple benchmark scenario where only one heavy neutrino state is taken into account. 

Thus we consider an effective Lagrangian in which we include only \emph{one} relatively light right handed Majorana neutrino $N$ as an observable degree of freedom. The effects of the new physics involving one heavy sterile neutrino and the SM fields are parameterized by a set of effective operators $\mathcal{O}_\mathcal{J}$ constructed with the standard model and the Majorana neutrino field and satisfying the $SU(2)_L \otimes U(1)_Y$ gauge symmetry \cite{Wudka:1999ax}. 

The effect of these operators is suppressed by inverse powers of the new physics scale $\Lambda$. The total Lagrangian is organized as follows:

\begin{eqnarray}\label{eq:Lagrangian}
\mathcal{L}=\mathcal{L}_{SM}+\sum_{n=5}^{\infty}\frac1{\Lambda^{n-4}}\sum_{\mathcal{J}} \alpha_{\mathcal{J}} \mathcal{O}_{\mathcal{J}}^{(n)}
\end{eqnarray}
where $n$ is the mass dimension of the operator $\mathcal{O}_{\mathcal{J}}^{(n)}$.

Note that we do not include the Type-I seesaw Lagrangian terms giving the Majorana and Yukawa terms for the sterile neutrinos. The dominating effects come from the lower dimension operators that can be generated at tree level in the unknown underlying renormalizable theory.

The dimension 5 operators were studied in detail in \cite{Aparici:2009fh}. These include the well known Weinberg operator $\mathcal{O}_{W}\sim (\bar{L}\tilde{\phi})(\phi^{\dagger}L^{c})$ \cite{Weinberg:1979sa} contributing to the light neutrino masses, and operators with the $N$: $\mathcal{O}_{N\phi}\sim (\bar{N}N^{c})(\phi^{\dagger} \phi)$ contributing to the $N$ Majorana masses and giving couplings of the heavy neutrinos to the Higgs (its phenomenology for the LHC has been studied very recently in \cite{Caputo:2017pit}), and an operator $\mathcal{O}^{(5)}_{NB}\sim (\bar{N}\sigma_{\mu \nu}N^{c}) B^{\mu \nu}$ inducing magnetic moments for the heavy neutrinos, which is identically zero if we include just one sterile neutrino $N$ in the theory \footnote{The effects of considering the $\mathcal{O}^{(5)}_{NB}$ operator were studied in \cite{Aparici:2009fh} for the case of 2 massive Majorana neutrinos $N_{1,2}$. The lighter $N_{1}$ must decay to SM particles via the mixing between light and heavy neutrinos. The channel $N_1\rightarrow \nu \gamma$ is also found to be dominant for $m_{N_{1}}<m_W$, and its possible discovery through the presence of a displaced photon vertex is mentioned. Our treatment coincides with the limit in which $N_{1,2}$ are mass-degenerate and the light-heavy mixing is taken to be zero.}. 

In the following, as the dimension 5 operators do not contribute to the studied processes -discarding the heavy-light neutrino mixings- we will only consider the contributions of the dimension 6 operators, following the treatment made in \cite{delAguila:2008ir}. We start with a rather general effective Lagrangian density for the interaction of right-handed Majorana neutrinos $N$ including dimension 6 operators.

The first operators subset includes those with scalar and vector bosons (SVB), 
\begin{eqnarray} \label{eq:Ope-SVB}
\mathcal{O}_{LN\phi}=(\phi^{\dag}\phi)(\bar L_i N \tilde{\phi}),
\;\; \mathcal{O}_{NN\phi}=i(\phi^{\dag}D_{\mu}\phi)(\bar N
\gamma^{\mu} N), \;\; \mathcal{O}_{Ne\phi}=i(\phi^T \epsilon D_{\mu}
\phi)(\bar N \gamma^{\mu} e_i)
\end{eqnarray}
and a second subset includes the baryon-number conserving four-fermion contact terms:  
\begin{eqnarray} \label{eq:Ope-4-f}
\mathcal{O}_{duNe}&=&(\bar d \gamma^{\mu} u)(\bar N \gamma_{\mu} l) , \;\; \mathcal{O}_{fNN}=(\bar f \gamma^{\mu}
f)(\bar N \gamma_{\mu}
N), \;\; \mathcal{O}_{LNLe}=(\bar L N)\epsilon (\bar L l),
\nonumber \\
\mathcal{O}_{LNQd}&=&(\bar L N) \epsilon (\bar Q
d), \;\; \mathcal{O}_{QuNL}=(\bar Q u)(\bar N L) , \;\; \mathcal{O}_{QNLd}=(\bar Q N)\epsilon (\bar L d), 
\end{eqnarray}
where $e_i$, $u_i$, $d_i$ and $L_i$, $Q_i$ denote, for the family
labeled $i$, the right handed $SU(2)$ singlet and the left-handed
$SU(2)$ doublets, respectively. Also $ \gamma^{\mu}$ and $\sigma^{\mu \nu}$ are the Dirac matrices, and $\epsilon=i\sigma^{2}$ is the antisymmetric symbol.

One can also consider operators generated at one-loop level in the underlying full theory, whose coefficients are naturally suppressed by a factor $1/16\pi^2$\cite{delAguila:2008ir, Arzt:1994gp}:
\begin{eqnarray}\label{eq:Ope-1-loop} 
\mathcal{O}_{ N B} = (\bar L \sigma^{\mu\nu} N) \tilde \phi B_{\mu\nu} , &&
\mathcal{O}_{ N W } = (\bar L \sigma^{\mu\nu} \tau^I N) \tilde \phi W_{\mu\nu}^I
\end{eqnarray}
Here $B_{\mu\nu}$ and $W_{\mu\nu}^I$ represent the $U(1)_{Y}$ and $SU(2)_{L}$ field strengths respectively. 

In this paper we study the well known ss-dilepton Majorana neutrino signal $pp\rightarrow  l_i^{+} l_j^{+} j j$ schematically depicted in Fig.\ref{fig:uD_Nl_ljj}, and the new $pp\rightarrow l_i^{+} \nu \gamma$ neutrino plus photon channel shown in Fig.\ref{fig:uD_Nl_lnug}. 
In order to obtain the cross sections for the above processes, we derive the effective Lagrangian terms involved in the calculations, taking the scalar doublet after spontaneous symmetry breaking as $\phi=\tvect{0}{\frac{v+h}{\sqrt{2}}}$, with $h$ being the Higgs field and $v$ its vacuum expectation value.

For the production vertex $I$ in Figs.\ref{fig:uD_Nl_ljj} and \ref{fig:uD_Nl_lnug}, and the decay vertex $II$ in Fig.\ref{fig:uD_Nl_ljj}, we have tree-level generated contributions from the effective Lagrangian coming from \eqref{eq:Ope-SVB}, related to the spontaneous symmetry breaking process, and $4$-fermion contributions from \eqref{eq:Ope-4-f}:
\begin{eqnarray}\label{eq:leff_tree}
\mathcal{L}^{tree-level}_{eff}&=&\frac{1}{\Lambda^2}\left\{- \frac{m_W v}{\sqrt{2}} 
\alpha^{(i)}_W
\; W^{\dag\; \mu} \; \overline N_R \gamma_{\mu} e_{R,i} + \alpha^{(i)}_{V_0} 
\bar d_{R,i} \gamma^{\mu} u_{R,i} \overline N_R \gamma_{\mu}
e_{R,i} + \right.
\nonumber
\\ &&
 \alpha^{(i)}_{S_1}(\bar
u_{L,i}u_{R,i}\overline N \nu_{L,i}+\bar d_{L,i}u_{R,i} \overline N e_{L,i})
 +
\alpha^{(i)}_{S_2} (\bar \nu_{L,i}N_R \bar d_{L,i}d_{R,i}-\bar
e_{L,i}N_R \bar u_{L,i}d_{R,i}) +
\nonumber
\\ &&
\left. \alpha^{(i)}_{S_3}(\bar u_{L,i}N_R
\bar e_{L,i}d_{R,i}-\bar d_{L,i}N_R \bar \nu_{L,i}d_{R,i})
  + h.c. \right\}
\end{eqnarray}
where the sum over the families $i$ is understood  and the constants
$\alpha^{(i)}_{\mathcal{J}}$ are associated to specific operators according to
\begin{eqnarray}\label{eq:alphas}
\alpha^{(i)}_W=\alpha^{(i)}_{Ne\phi},\;
\alpha^{(i)}_{V_0}=\alpha^{(i)}_{duNe},\;\;
\alpha^{(i)}_{S_1}&=&\alpha^{(i)}_{QuNL},\;
\alpha^{(i)}_{S_2}=\alpha^{(i)}_{LNQd},\;\;
\alpha^{(i)}_{S_3}=\alpha^{(i)}_{QNLd}~.\;
\end{eqnarray}

For the $N\rightarrow \nu \gamma$ decay vertex $II$ in Fig.\ref{fig:uD_Nl_lnug}, the considered Lagrangian terms are generated by one-loop level tensorial operators: 
\begin{eqnarray}\label{eq:leff_1loop_L}
\mathcal{L}_{eff}^{1-loop}&=& \frac{-i\sqrt{2} v}{\Lambda^2} {(\alpha_{NB}^{(i)}c_W + 
 \alpha_{NW}^{(i)}s_W)  (P^{(A)}_{\mu} ~\bar \nu_{L,i} \sigma^{\mu\nu}N_R~ A_{\nu})}.
\end{eqnarray}
where $-P^{(A)}$ is the 4-momentum of the outgoing photon and a sum over the family index $i$ is understood again. 
The coupling constants $\alpha^{(i)}_{NB}$ and $\alpha^{(i)}_{NW}$ correspond respectively to the operators in \eqref{eq:Ope-1-loop}. The total decay width of the $N$ ($\Gamma_{N}$) is calculated in \cite{Duarte:2016miz}, and the complete dimension 6 effective Lagrangian is presented in an appendix in that work. 

The effective operators in \eqref{eq:Ope-SVB} and \eqref{eq:Ope-4-f} cover a wide variety of new physics models, as extended scalar and gauge sectors, vector and scalar leptoquarks, heavy fermions, etc. For example, the four-fermion contact operators $\mathcal{O}_{duNe}\equiv \mathcal{O}_{V_{0}}$ and $\mathcal{O}_{QNLd}\equiv \mathcal{O}_{S_{3}}$ have been studied recently as a parameterization of the minimal Left-Right Symmetric Model (LRSM) in recasts of LHC searches for the same-sign dilepton signal \cite{Ruiz:2017nip}. 

The effective operators above can be classified by their Dirac-Lorentz structure into {\it{scalar}}, {\it{vectorial}} and {\it{tensorial}}. The scalar and vectorial operators contributing to the studied processes are those appearing in \eqref{eq:leff_tree} with couplings named $\alpha_{S_{1,~2,~3}}$ and $\alpha_{W,~ V_{0}}$ respectively. They play a role in the production and decay vertices for the $pp \rightarrow l_i^{+} l_j^{+} +  2 ~jets$ process, and only in the production vertex for the $l_{i}^+ \nu \gamma$ channel. The one-loop tensorial operators in \eqref{eq:leff_1loop_L} give a magnetic moment for Majorana neutrinos, which drives the $N\rightarrow \nu \gamma$ decay.

\subsection{Effective coupling bounds summary}\label{sect:bounds}

The couplings $\alpha^{i}_{\mathcal{J}}$ of the different operators $\mathcal{O}_{\mathcal{J}}$ in the effective Lagrangian \eqref{eq:Lagrangian} can be bounded by exploiting the current existing experimental constraints on right-handed sterile Majorana neutrinos, which are generally imposed on the parameters representing the light-heavy neutrinos mixing parameters in seesaw models. 

Recent reviews \cite{deGouvea:2015euy, Deppisch:2015qwa, Antusch:2015mia, Das:2014jxa, Fernandez-Martinez:2016lgt} summarize in general phenomenological approaches the existing experimental bounds, considering low scale minimal seesaw models, parameterized by a single heavy neutrino mass scale $M_{N}$ and a light-heavy mixing $U_{lN}$, with $l$ indicating the lepton flavor. In previous works \cite{Duarte:2016miz, Duarte:2015iba} we have presented in detail the way in which we take into account existing constraints on processes like neutrino-less double beta decay ($0\nu\beta\beta$), electroweak precision data (EWPD), LNV rare meson decays as well as direct collider searches, including $Z$ decays. We refer the reader to those papers for a detailed discussion. In order to put reliable bounds for the effective couplings, we take into account existing experimental constraints on sterile-active neutrino mixings, relating the $U_{lN}$ mixings in Type-I seesaw models \cite{Atre:2009rg, delAguila:2007qnc} with our effective couplings $\alpha_{\mathcal{J}}$ in \eqref{eq:Lagrangian} by the relation 
\begin{equation}
U^2_{lN}\simeq \left(\frac{\alpha_{\mathcal{J}}~ v^2}{2\Lambda^2}\right)^2, 
\end{equation} 
for the operators $\mathcal{O}_{\mathcal{J}}$ contributing to each process imposing bounds on the mixings $U^2_{lN}$. 

For the couplings involving the first fermion family the most stringent are the $0\nu\beta\beta$-decay bounds obtained by the KamLAND-Zen collaboration \cite{KamLAND-Zen:2016pfg}. Following the treatment made in \cite{Mohapatra:1998ye, deGouvea:2015euy, Duarte:2016miz}, they give us an upper limit  $\alpha^{bound}_{0\nu\beta\beta} \leq 3.2 \times 10^{-2} \left(\frac{m_N}{100 ~GeV}\right)^{1/2}$, where the new physics scale is taken to be $\Lambda=1~TeV$ (here and in the following) \footnote{The new physics scale $\Lambda=1~TeV$ is taken as an illustration. The scale can be changed to any other scale $\Lambda^{\prime}$ considering $\alpha^{\prime}_{\mathcal{J}}=(\frac{\Lambda^{\prime}}{\Lambda})^{2} \alpha_{\mathcal{J}}$. }.
Concerning the second fermion family, for sterile neutrino masses $2~GeV\lesssim m_{N}\lesssim 10~GeV$ the upper limits come from the DELPHI collaboration \cite{Abreu:1996pa}. Considering $\Omega_{l l'} = U_{l N} U_{l' N}$ as in  \cite{delAguila:2007qnc}, we obtain the bound $\alpha^{bound}_{DELPHI} \lesssim 2.3$. 
The Belle \cite{Liventsev:2013zz} and LHCb \cite{Aaij:2014aba} collaborations also find competitive upper limits in the $2~GeV\lesssim m_{N}\lesssim 5~GeV$ region. These results though depend heavily on the  considered decay modes for the sterile $N$ \cite{Shuve:2016muy, Cvetic:2016fbv, Milanes:2016rzr, Mandal:2016hpr, Dib:2014iga}. The bound from Belle is still the most stringent, giving a value $\alpha^{bound}_{Belle} \lesssim 0.3$.
For higher masses, in the range $m_W\lesssim m_N$ the upper limits come from EWPD as radiative lepton flavor violating (LFV) decays as $\mu\rightarrow e \gamma$ \cite{Tommasini:1995ii, delAguila:2008pw,  deGouvea:2015euy, Fernandez-Martinez:2016lgt} giving a bound $\alpha^{bound}_{EWPD} \leq 0.32$. 

Besides the different effective couplings are bounded in general by different experiments, in order to simplify the discussion, for the numerical evaluation of the cross sections and decay widths in this work we consider two different sets of numerical values for them. In the set we call set 0 ({\bf s0}), we take the couplings associated to the operators that contribute to the $0\nu\beta\beta$-decay for the first family as restricted by the corresponding bound $\alpha^{bound}_{0\nu\beta\beta}$ \footnote{These are the couplings in \eqref{eq:alphas}, for the first family ($i=1$). }, and we fix the other constants to the value  $\alpha^{bound }\leq 0.3$. For the 1-loop generated operators we consider the coupling constant as $1/(16 \pi^2)$ times the corresponding tree-level coupling: $\alpha^{1-loop}=\alpha^{tree}/(16 \pi^2)$. Thus, for example, for the operator $\mathcal{O}_{NW}$, which contributes to $0\nu\beta\beta$ we have 
\begin{equation*}
\alpha^{(1)}_{NW} = \frac{1}{16\pi^2} \alpha^{bound}_{0\nu\beta\beta}.
\end{equation*}
The set we call set 1 ({\bf s1}) takes all the effective couplings to be equal to $\alpha^{bound}_{0\nu\beta\beta}$. As an illustration, for a Majorana mass $m_N= 5 ~GeV$ the $0\nu\beta\beta$ bound takes the value $\alpha^{bound}_{0\nu\beta\beta}(m_N= 5 ~GeV)=4.5\times 10^{-3}$. Thus set 0 ({\bf s0}) puts higher numerical values for the couplings than set 1 ({\bf s1}).

\begin{table}[t]
 \centering
 \begin{tabular}{c c c r}
\firsthline
Set ~~& $\alpha_{0\nu\beta\beta}$~~ & $\alpha_{others}$ & $\alpha_{1-loop}$\\
\hline
 {\bf s0} & $\alpha^{bound}_{0\nu\beta\beta}$  & $0.3$  &  $1.9\times 10^{-3}$\\
 {\bf s1} & $\alpha^{bound}_{0\nu\beta\beta}$  & $\alpha^{bound}_{0\nu\beta\beta}$ &  ~~$ (0.02) \times 10^{-2} \left(\frac{m_N}{100 ~GeV}\right)^{1/2}$\\
\lasthline
 \end{tabular}
\caption{Effective couplings values in the considered sets.
 Here $\alpha^{bound}_{0\nu\beta\beta}= 3.2 \times 10^{-2} 
\left(\frac{m_N}{100 ~GeV}\right)^{1/2}$, $\Lambda=1 ~TeV$. }\label{tab:alpha-sets}
\end{table}

\subsection{Low $m_N$ kinematic features}\label{sect:low_mN}

As we mentioned in the introduction, Majorana neutrinos with masses of a few $GeV$ are found to be long-lived neutral particles that could be searched for in the LHC with displaced vertices techniques. In this paper we exploit this long decay length in order to search for the production of Majorana neutrinos.

In higher $m_N$ regions, the same-sign dilepton ($l^{+}l^{+} jj$) signal has been thoroughly studied in hadron colliders, in the effective framework we consider here \cite{delAguila:2008ir}, and searched for in the LHC \cite{Aad:2011vj, ATLAS:2012ak} for $m_N$ above $100~GeV$. Up to now, all LHC searches for heavy Majorana neutrinos \cite{Khachatryan:2016olu, Khachatryan:2015gha, Aad:2015xaa, Aad:2011vj, ATLAS:2012ak} in these final states are performed without considering the possibility of a sufficiently long heavy neutrino lifetime causing the decay vertex to be displaced from the production point. 

The average decay length $\ell_{N}$ of the Majorana neutrino and its flight direction in the lab frame can be obtained for each event from their simulated momenta $k_{N}$:  
\begin{equation}\label{eq:l_N}
\ell_{N} =  \tau_{N} \gamma \beta_{N}= \frac{\left(\left(E_{N}/m_N\right)^2-1\right)^{1/2}}{\Gamma_{N}}
\end{equation}
where $E^{2}_{N}=|\vec{k}_{N}|^2+m^{2}_{N}$ and $\Gamma_{N}=1/\tau_{N}$ is the decay width. The distance $\ell$ traveled by each Majorana neutrino $N$ in the lab frame is randomly sampled according to an exponential distribution $f_{dec}(\ell, \ell_{N})= \frac{1}{\ell_{N}}{e^{-\frac{\ell}{\ell_{N}}}}$, and the $N$ displacement vector is 
\begin{equation}\label{eq:L_N}
 \vec{L}_{N} = \frac{\vec{k}_{N}}{|\vec{k}_{N}|} \ell.
\end{equation}
As the Majorana neutrinos can travel a macroscopic distance before decaying, we consider only those events with the decay occurring inside the detector. The probability for the $N$ to decay inside a detector of size $L_{D}$ is $P_{DI}= (1- e^{-\frac{L_{D}}{\ell_{N}}})$. 

For each considered process in this work, we expect a number of signal events which depends on the theoretical model parameters $m_N$ and the effective couplings set $\alpha_{set}$ (as defined in sec.\ref{sect:bounds}) through the scattering cross section $\sigma_{proc}(m_N,\alpha_{set})$, on the integrated luminosity $\mathcal{L}$, and on the size of the detector $L_{D}$:
%
\begin{equation}\label{eq:Nevents}
\mathcal{N}(m_{N}, \alpha_{set}, \mathcal{L}) = \mathcal{L} ~\sigma_{proc}(m_N,\alpha_{set}) ~P_{DI}. 
\end{equation}
%

The measurable decay length of the few-$GeV$ mass Majorana neutrino will be exploited in this paper considering two distinct observables: for the ss-dilepton signal, we find the vertex displacement can be measured by obtaining the distance between the traces of both outgoing leptons, a prompt one coming from the $N$ production vertex, and the other in the decay vertex. For the neutrino plus photon channel, we consider the distance between the prompt outgoing lepton and the displaced photon traces, and also the non-pointing photon observables defined in recent new physics searches in ATLAS \cite{Aad:2013oua, Aad:2014gfa}.  

\begin{figure*}[h]
\begin{center}
\subfloat[Anti-muon channel]{\label{fig:sig_mu}\includegraphics[width=0.5\textwidth]{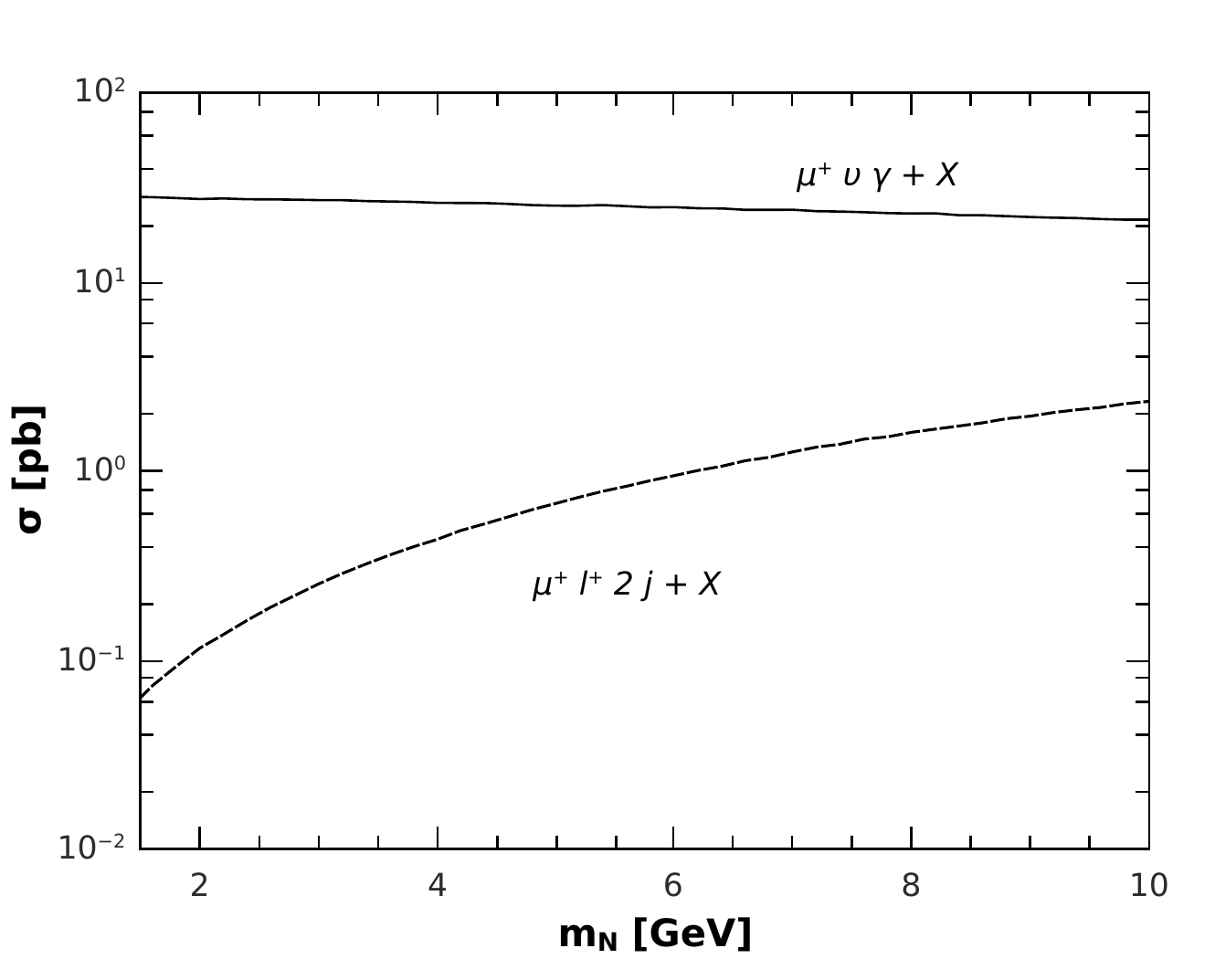}}
~
\subfloat[Positron channel]{\label{fig:sig_e}\includegraphics[width=0.5\textwidth]{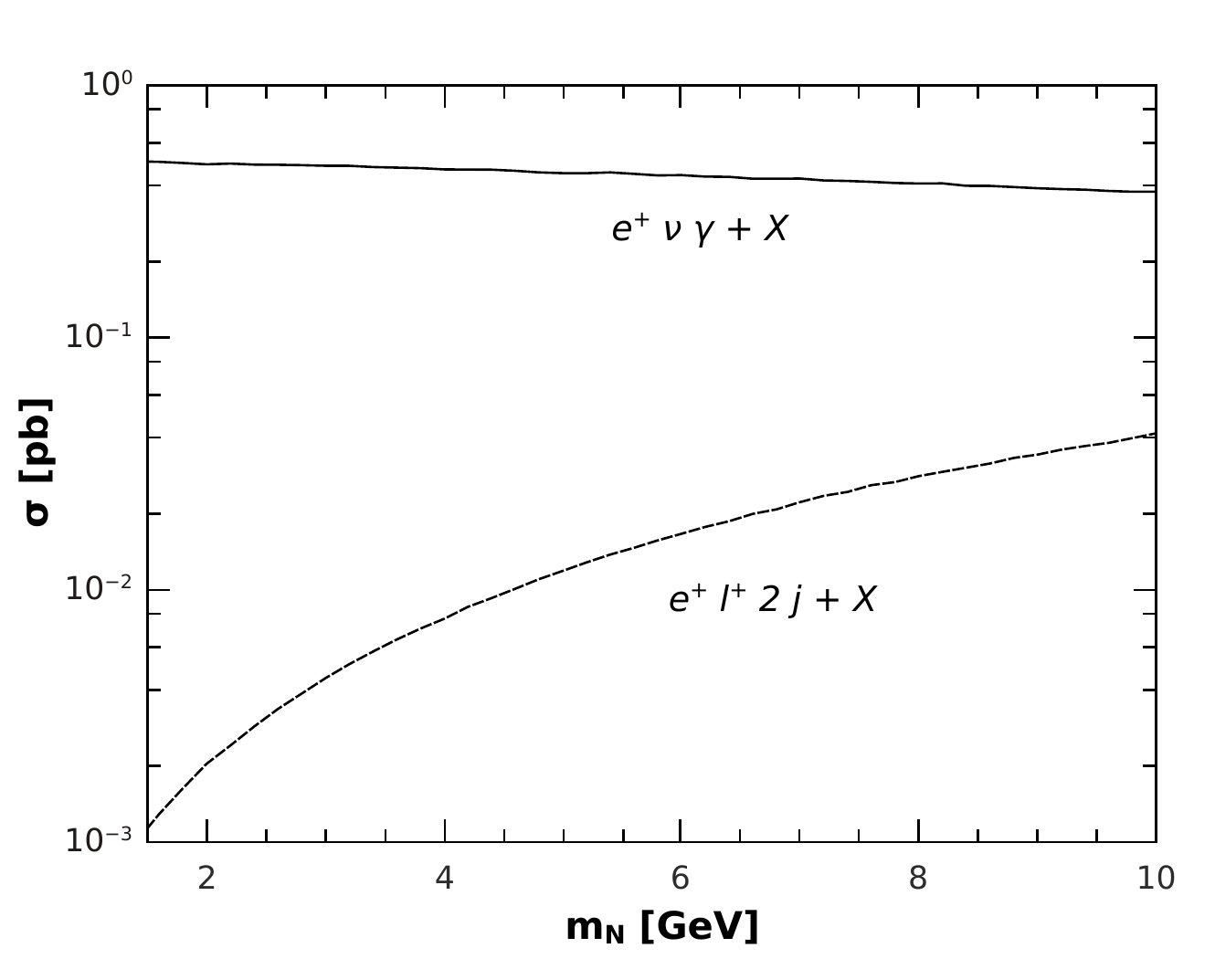}}
\caption{\label{fig:sigmas} Total cross sections for the $p p \rightarrow l^{+} \nu \gamma + X$ and $p p \rightarrow l^{+}l^{+}+2~j+ X$ processes, for the effective couplings set 0 ({\bf s0}). $\Lambda= 1 ~TeV$. All the effective operators in \eqref{eq:leff_tree} and \eqref{eq:leff_1loop_L} are included.}
\end{center}
\end{figure*}

The cross sections for the $l^{+}l^{+} jj$ and the $l^{+} \nu \gamma$ final states are shown in Fig.\ref{fig:sigmas}, for the prompt anti-muon (\ref{fig:sig_mu}) and prompt positron (\ref{fig:sig_e}) channels. Here we take into account all the scalar, vectorial and tensorial operators contribution (in eqs. \eqref{eq:Ope-SVB}, \eqref{eq:Ope-4-f} and \eqref{eq:Ope-1-loop}), and set the numerical values of the effective couplings according to set 0 ({\bf s0}) in Tab.\ref{tab:alpha-sets}. In this plots we do not take into account the probability factor $P_{DI}$ in \eqref{eq:Nevents}.

It can be clearly seen in Fig.\ref{fig:sigmas} that in the effective framework we are working with, the dominant decay channel in the few-$GeV$ $m_N$ mass region is the $N\rightarrow \nu \gamma$. 
Considering an on shell $N$ production, the cross section for the ss-dilepton process can be written as the product of the production cross section by the decay channel branching ratio: $\sigma (pp\rightarrow ll jj) \simeq \sigma (u \bar{d} \rightarrow l N) Br(N\rightarrow l jj)$. 

As the radiative $N$-decay channel completely dominates the low mass region and is driven by tensorial one-loop operators, for the study of the LNV same-sign dileptons plus jets signal, we assume $\alpha_{1-loop}=0$ in order to study the ss-dilepton channel, which is not observable if we have a $1-loop$ effective operators contribution.

In Fig.\ref{fig:sigmas} one can also appreciate the diminishing effect of the neutrino-less double beta decay bound $\alpha^{bound}_{0\nu\beta\beta}$ value on the $l^{+}=e^{+}$ channel: the cross section for the second family $l^{+}=\mu^{+}$ is appreciably higher, due to the more relaxed bound $\alpha^{bound}=0.3$ considered in set 0 ({\bf s0}).  

\section{Same sign dilepton signal}\label{sect:ss_dilepton}

\begin{figure*}[h]
\begin{center}
\includegraphics[width=0.5\textwidth]{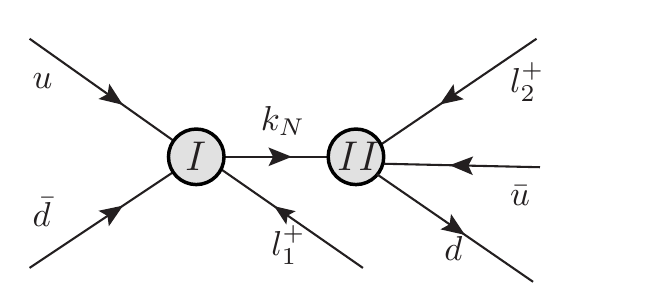}
\caption{\label{fig:uD_Nl_ljj}  Diagrams contributing to the process $pp\rightarrow l^{+}l^{+} j j$.}
\end{center}
\end{figure*}

Using the effective Lagrangian in \eqref{eq:leff_tree} we calculate the cross section for the production of the heavy neutrino according to the parton hard processes shown in Fig.\ref{fig:uD_Nl_ljj}.

The corresponding amplitude can be written as the product of the production and decay processes:
\begin{eqnarray}\label{eq:Mssdilepton}
\mathcal{M}&&=\frac{-i m_N}{\Lambda^2} P_N(k_N^2)\left\{ \Pi_W(q_2^2)\alpha_W^{(i)} \, \bar u(l_u)\gamma_{\nu} P_R v(l_d) \bar u(l_2)\gamma^{\nu}
+ \alpha_{V_0}^{(i)} \, \bar u\gamma_{\nu}P_L v(l_d) \bar u(l_2) \gamma^{\nu} + \right.
\nonumber \\
&& \left. \alpha_{S_1}^{(i)} \, \bar u(l_u) P_R v(l_d) \bar u(l_2)-\alpha_{S_2}^{(i)} \, \bar u(l_u) P_L v(l_d) \bar u(l_2) + \right.
\nonumber \\
&& \left. \alpha_{S_3}^{(i)} \, \bar u(l_2) P_L v(l_d) \bar u(l_u) \right\} \otimes 
\left\{-\Pi_W(q_1^2) \alpha_W^{(i)} \, \gamma_{\mu} P_R v(l_1) \bar v(p_d)\gamma^{\mu} P_L u(p_u)+ \right.
\nonumber \\
&& \left. \alpha_{V_0}^{(i)} \, \gamma_{\mu} P_R v(l_1) \bar(p_d) \gamma^{\mu} P_R u(p_u) + 
\alpha_{S_1}^{(i)} \, P_L v(l_1) \bar v(p_d) P_R u(p_u)- \right.
\nonumber \\
&& \left. \alpha_{S_2}^{(i)} \, P_L v(l_1) \bar v(p_d) P_L u(p_u)+
\alpha_{S_3}^{(i)} \, P_L u(p_u) \bar v(p_d) P_L v(k_1)\right\},
\end{eqnarray}
where $l_1$ and $l_2$ are the 4-moments of the final leptons, $p_u$ and $p_d$ are the 4-moments of the initial quarks and, $l_u$ and $l_d$ stand for the 4-moments of the final quarks. The $W$ propagator is $\Pi_W$, $k_N$ is the 4-moment of the intermediate Majorana neutrino, and $P_{N}$ its propagator.

Taking the center of mass energy $\sqrt{s}=14$ T$e$V, $\hat{\sigma}$ 
and $\hat{s}$ to be the parton level scattering cross section and squared 
center of mass energy, and $x_1$ and $x_2$ the usual deep inelastic scaling variables, 
we write the cross section as
\begin{eqnarray}
\sigma(pp\rightarrow l^+ \, l^+ \, +\,  2 jets)=\int_{x_m}^1 \int_{x_m/x_1}^1 dx_1 dx_2 
( f_u(x_1) f_{\bar d}(x_2) + f_u(x_2) f_{\bar d}(x_1) )  \hat \sigma_i(x_1 x_2 s)
\end{eqnarray}
where the minimum value for $x_1$ and $x_2$ is $x_m = m_N^2/s$ and the function $f_q(x)$ represents the $q(x)$ parton distribution function (PDF). In our numerical simulations we have used the CTEQ functions \cite{Pumplin:2002vw}. 

\subsubsection{Backgrounds}

As the considered signal is a LNV process, it is strictly forbidden in the SM, and the background always involves additional light neutrinos in the final state that escape the detectors as missing energy. The SM backgrounds have been extensively studied in the literature (see \cite{Atre:2009rg, delAguila:2007qnc, Alva:2014gxa, Chakdar:2016adj, Bhattacharya:2015vja}) and in recent experimental searches \cite{Khachatryan:2015gha, Khachatryan:2016olu}.

The background for these searches can be classified into prompt leptons, charge-flip opposite-sign dileptons and misidentified leptons.
The first are SM events resulting in two genuine same-sign leptons, including diboson production ($WZ$, $ZZ$) which gives ss-dileptons when both $Z$ and $W$ decay leptonically and one lepton from $Z$ decay cannot be isolated or is missed out of the detector coverage, $t\bar{t}$-plus boson ($t\bar{t}W$, $t\bar{t}W$, $t\bar{t}Z$) processes contributing to the signal when the tops decay hadronically and the bosons decay leptonically, triboson ($W^{\pm}W^{\pm}W^{\mp}$), double $W$-sstrahlung ($W W j j$) and double parton scattering ($q q'\rightarrow W$). 
The charge-flip events originate in opposite-sign dileptons signals in which an electron undergoes bremsstrahlung in the tracker volume and the associated photon decays into en $e^{+}e^{-}$ pair, and the opposite sign electron is misidentified as the primary electron if it carries a large fraction of the original electron's energy. This effect is negligible for muons. 
The last group includes objects misidentified as prompt leptons, originated in $B$-hadrons decays, light quark or gluon jets and photon conversions, and mainly $t\bar{t}$ in which a top quark yields a prompt isolated lepton ($t\rightarrow W b \rightarrow l \nu b$) and the other same charge lepton arises from $b$ quark decay or a jet misidentified as an isolated prompt lepton. This is the dominant background for the $m_N < m_W$ region \cite{Khachatryan:2016olu, delAguila:2007qnc}, and it cannot be easily eliminated in the few-$GeV$ $m_N$ region, as the cuts imposed on the final leptons $p_T$ in order to reject those coming from $b$ decays can also affect the signal: the tracks have not enough momentum to pass the cuts. This channel can be viable with relaxed kinematic criteria, but as stressed in \cite{Izaguirre:2015pga}, we lack the tools to simulate this background, as such a signature would need to be performed with a data-driven estimation, as is done in \cite{Khachatryan:2016olu}.   

In the range of $N$ masses we study, it is produced with a significant boost, and its products are typically collimated. However, distributions on the invariant mass of the $l_2^{+}jj$ system $M(l_2^{+}jj)\sim m_N$ could help to distinguish the signal \cite{Dib:2016wge, Dib:2015oka}.

Taking into account the vertex displacement can help background reduction, and that is why we study this very interesting feature for the low mass region. Recent LHC searches exploiting the long-lived particles displaced vertex feature in final states with charged leptons and jets \cite{Khachatryan:2014mea, CMS:2014hka, Aad:2015rba, Aad:2015uaa} originally intended for SUSY particles \cite{Evans:2016zau} are currently being used for studying models with right handed massive neutrinos \cite{Batell:2016zod}. 

\subsection{Numerical results}

In order to make a preliminary survey we study the same-sign dileptons plus jets channel making a parton level Monte Carlo simulation with the RAMBO \cite{Kleiss:1985gy} routine, for the LHC with $\sqrt{s}=14 ~TeV$. The effective new physics energy scale $\Lambda$ in \eqref{eq:Lagrangian} is taken to be $\Lambda = 1 \; TeV$, and the $CM$ energy of the hard processes $\sqrt{\hat{s}}\leq \Lambda$, to ensure the validity of the effective Lagrangian approach.  Throughout all this section, we consider events with $|\eta|<1.83$ for all the final state particles \footnote{This is done just to ensure the numerical stability of our Monte Carlo simulations.}. We do not introduce any cuts on transverse momentum, or track isolation. The numerical values for the effective couplings are taken in the set 0 ({\bf s0}) (Tab.\ref{tab:alpha-sets}). We present our numerical results summing over the final leptons possibilities ($l_1^{+} l_2^{+}= e\mu$, $\mu e$, $e e$, $\mu \mu$).

\begin{figure*}[h!]
\centering
\subfloat[ \scriptsize Schematic representation of the distance $L^{l^{+}l^{+}}$ between the final leptons directions]{\label{fig:ddv_sketch} \includegraphics[totalheight=6 cm]{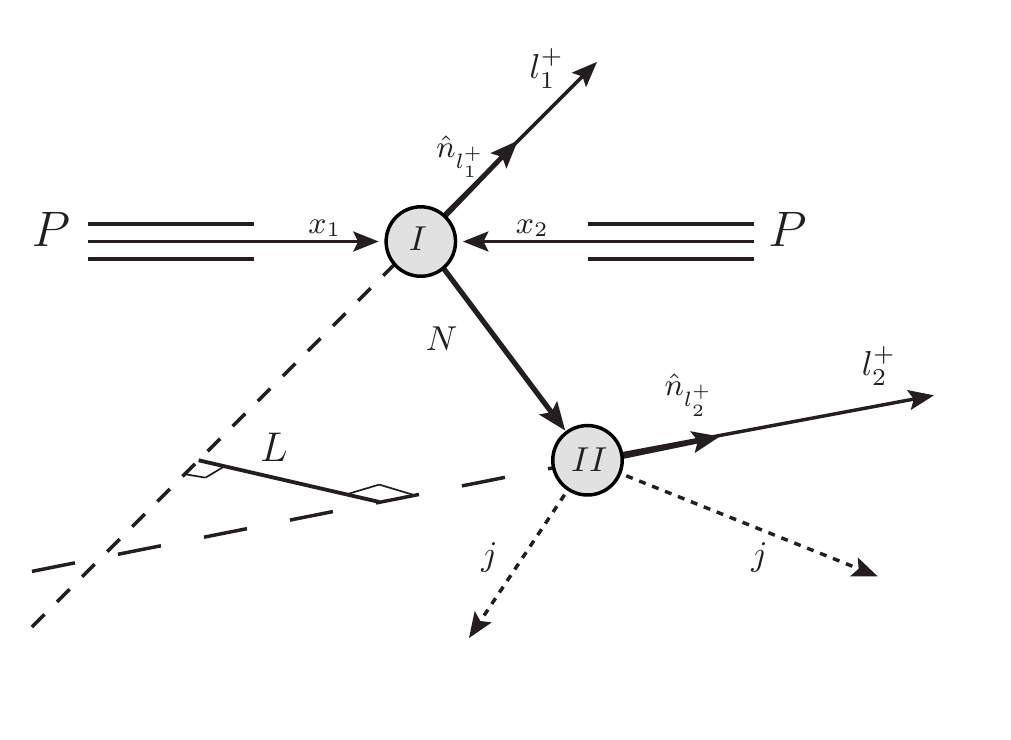}}
~
\subfloat[\scriptsize  Differential $L^{l^{+}l^{+}}$ distribution for different $m_N$ values. Scalar and vectorial operators are included, set 0 ({\bf s0}).]{\label{fig:dist_L} \includegraphics[totalheight=6 cm]{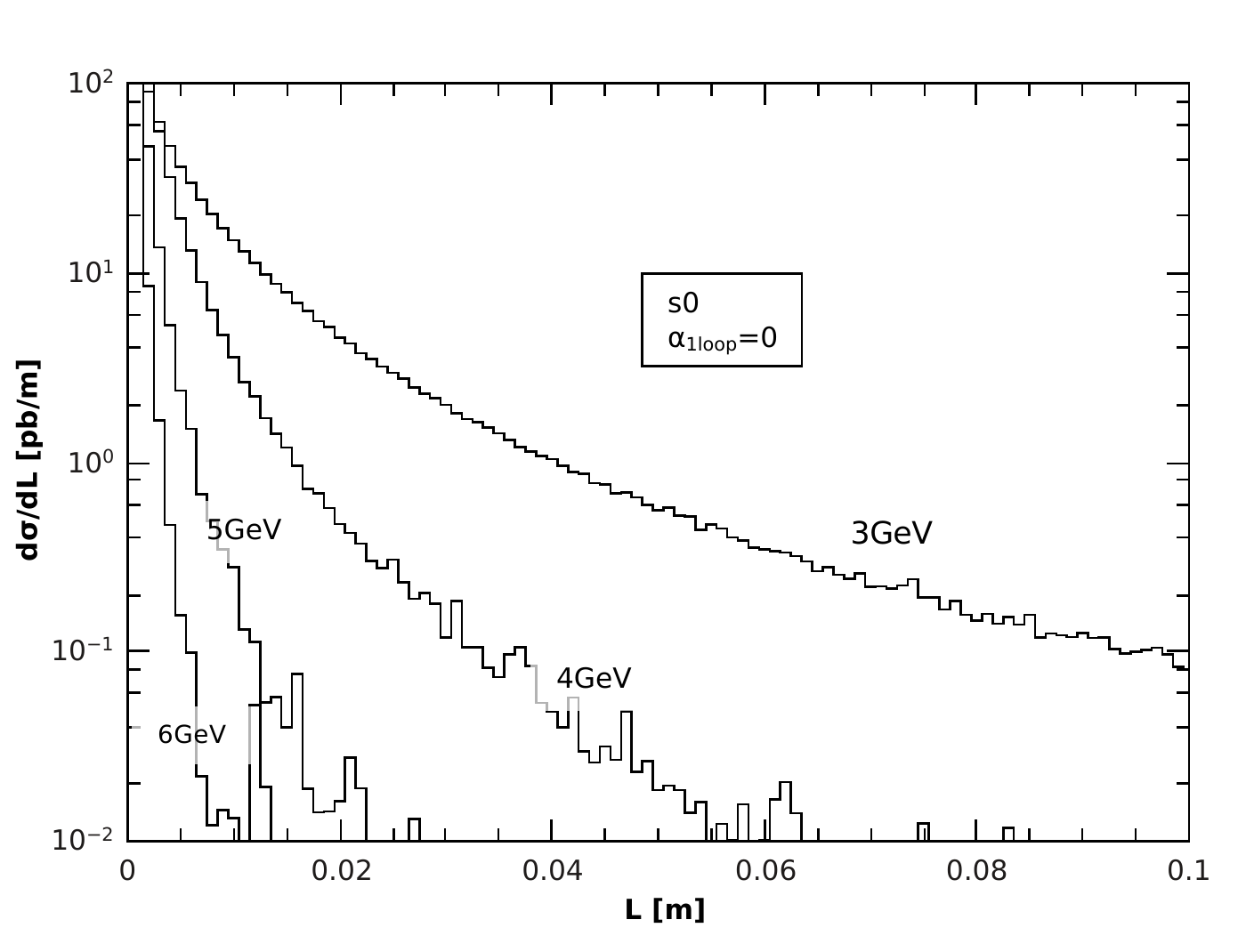}}
 \caption{\label{fig:L_ddv} $L^{l^{+}l^{+}}$ definition and differential distribution.}
\end{figure*}

In our parton level MC-simulation, the distance between the straight line containing the flight direction of the prompt lepton produced in the primary vertex and the one for the displaced secondary lepton can be found as the length of the segment orthogonal to and crossing both traces: $\hat{n}_{l_1^{+}}$ and $\hat{n}_{l_2^{+}}$ in Fig. \ref{fig:ddv_sketch}. This line has direction $\hat{n}= \hat{n}_{l_1^{+}}\times \hat{n}_{l_2^{+}}/|\hat{n}_{l_1^{+}}\times \hat{n}_{l_2^{+}}|$ and the distance $L^{l^{+}l^{+}}$ can be calculated as $L^{l^{+}l^{+}}= \hat{n}\cdot \vec{L}_{N}$, with the definition given in \eqref{eq:L_N}.  

We calculate the distance between the traces of the charged leptons, which, almost always, is different from zero since the vertices are displaced. A sketch in Fig. \ref{fig:ddv_sketch} shows the $L^{l^{+}l^{+}}$ distance definition. The distribution of the signal distances between the lepton's fly directions is shown in Fig.\ref{fig:dist_L} for various few-$GeV$ $m_N$ masses. 

In order to reject the prompt ss-dilepton background, we set this cut up to $L^{l^{+}l^{+}}=1 ~mm$, which is approximately the precision of the detector, discarding events for which the distance between the traces of the final leptons is less than $1 ~mm$. In this way, we conveniently eliminate the background, since the distance between the traces would be zero if this final state (up to additional non-detectable neutrinos) comes from standard interactions.

\subsubsection{Leptons Asymmetry}\label{sect:lepton_assym}

Our aim in this section is to investigate if the contribution to the ss-dilepton signal from scalar and vectorial effective operators can be disentangled. This gives us a first estimate to know to what extent this kind of search is worth to be done. 

In view of the discussion given in Sect.\ref{sect:low_mN}, we study the few-$GeV$ $m_N$ region for the ss-dilepton signal taking into account the separate effects of the scalar and vectorial operators, and considering the tensorial operators contribution not to be present. 

To measure the effects from the scalar operators we set the coupling constants corresponding to the vector operators $\alpha_W$ and $\alpha_{V_0}$ equal to zero, and set the value of the scalar operators $\alpha^{(i)}_{S_{1,2,3}}$ in \eqref{eq:Mssdilepton} to the values of the bounds presented in Tab.\ref{tab:alpha-sets} for the set 0 ({\bf s0}). Similarly, to study the contribution from the vectorial effective operators we set the couplings $\alpha_{S_{1,2,3}}$ equal to zero, and take $\alpha_W=\alpha_{V_0}= \alpha^{bound}=0.3$ ({\bf s0}). 

\begin{figure*}[h!]
\centering
\subfloat[Vectorial]{\label{fig:dvV}\includegraphics[width=0.46\textwidth]{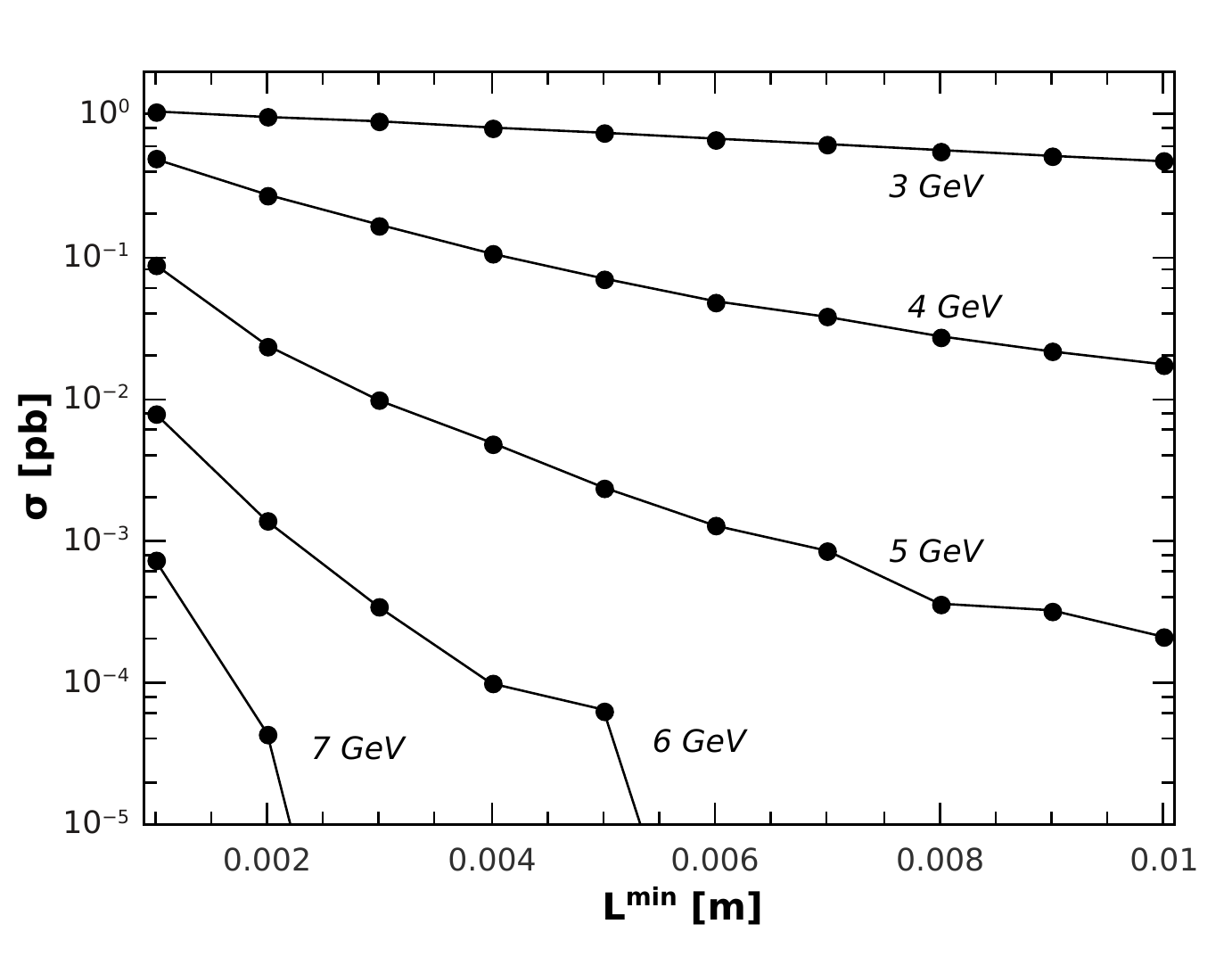}}
~
\subfloat[Scalar]{\label{fig:dvS}\includegraphics[width=0.5\textwidth]{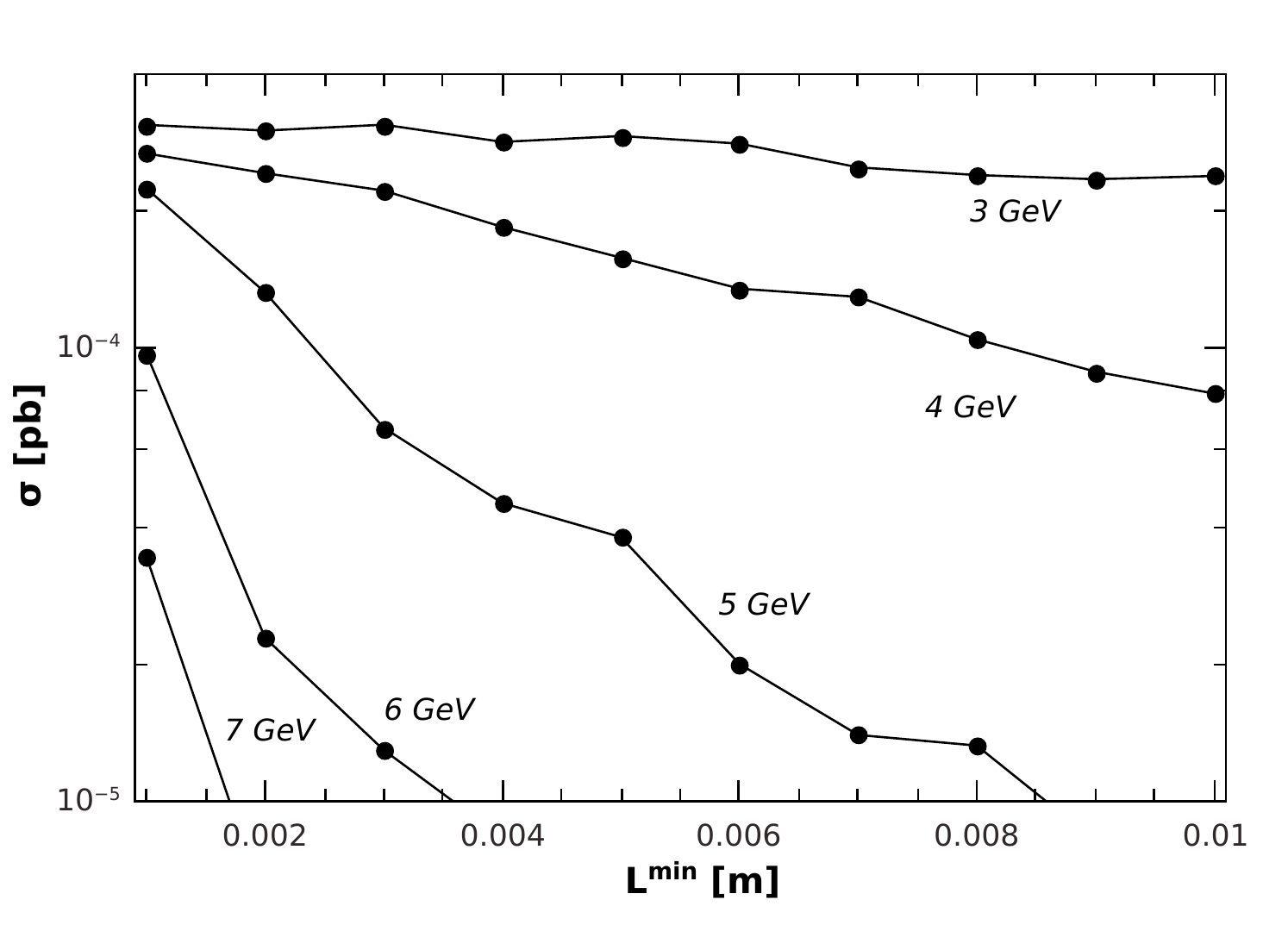}}
 \caption{\label{fig:dvLmin} Contribution of the vectorial and scalar operators to the cross section for different $m_N$ values as a function of the cut in $L^{l^{+}l^{+}}_{min}$, for the effective couplings set 0 ({\bf s0}).}
\end{figure*}

Once the displacement between the Majorana neutrino production and decay vertices is used to reduce the few-$GeV$ $m_N$ region background, we can test the separated contributions of the vectorial and scalar operators to the process, in order to obtain information about the kind of new physics behind these effects.

In Fig.\ref{fig:dvLmin} we show the contribution to the cross section of the vectorial (Fig.\ref{fig:dvV}) and scalar (Fig.\ref{fig:dvS}) operators for different $m_N$ mass values as a function of the implemented cut in the minimum distance between the lepton's fly directions $L^{l^{+}l^{+}}_{min}$.  
We find the vectorial operators give a greater cross section for the $pp\rightarrow l^+ \, l^+ \, +\,  2 jets$ process. This happens because for $m_N < m_W$ the vectorial $\alpha_{W}$ term in the amplitude \eqref{eq:Mssdilepton} leads to an s-channel $W$ resonance in the production vertex $I$, which is not present in the scalar-only case. It can be seen in Fig.\ref{fig:dvS} that for the applied cut on $L^{l^{+}l^{+}}_{min}=1~mm$ the cross section for $m_{N}=5 ~GeV$ is $\sigma_{scalar} \geq 2.1 \times 10^{-4} ~pb$. Considering an integrated luminosity of $100~fb^{-1}$, this gives $21$ signal events. In the case of vectorial operators (\ref{fig:dvV}) the number of signal events in the same case is $80000$. 

With the momentum vectors of the prompt lepton $l^{+}_{1}$ and the secondary lepton $l_{2}^{+}$, as shown in Fig. \ref{fig:ddv_sketch}, we can obtain the distribution in the angle $\theta$ between their tracks in the laboratory frame. This is done taking the scalar product of the unit vectors giving the directions of their tracks: $cos(\theta)= \hat{n}_{l_1^{+}} \cdot \hat{n}_{l_2^{+}}$.

\begin{figure*}
\begin{center}
\includegraphics[width=0.8\textwidth]{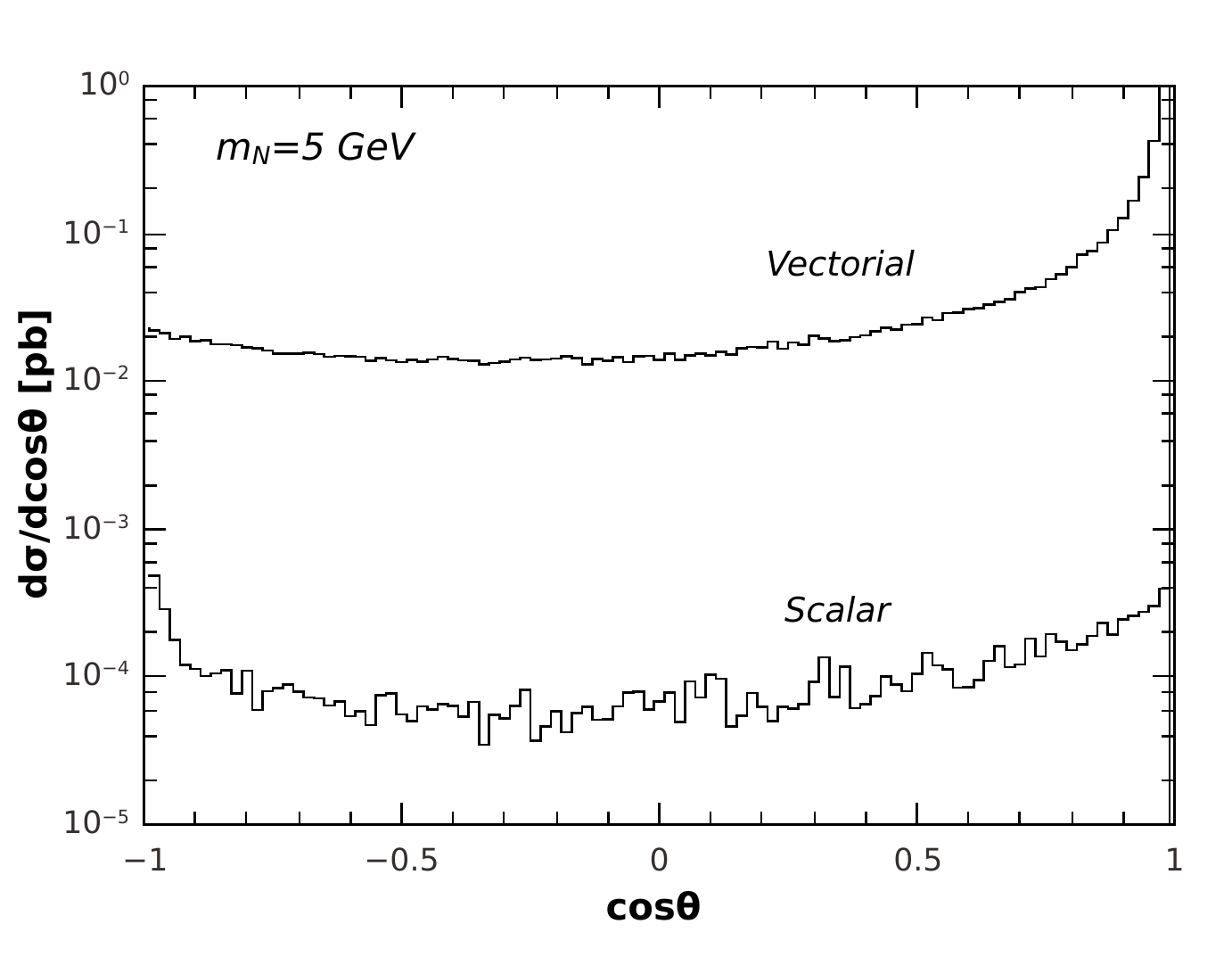}
\caption{\label{fig:asidv5n} Angular distribution of the final leptons ($l_1^{+} l_2^{+}= e\mu$, $\mu e$, $e e$, $\mu \mu$) for the vectorial and scalar operators. Here we take $m_N=5~ GeV$ and set 0 ({\bf s0}).}
\end{center}
\end{figure*}

An asymmetry $A^{l^+l^+}_{FB}$ can be constructed considering the difference between the number of events in the forward hemisphere $N_+$, with an angle $\theta$ in the interval $0\leq \theta \leq \pi/2$ -giving positive values for $cos(\theta)$- and the number of events in the backward hemisphere $N_-$, with $\pi/2\leq \theta \leq \pi$ and negative $cos(\theta)$:  
%
%
\begin{eqnarray}
A^{l^+l^+}_{FB}=\frac{N_+ \; - \; N_-}{N_+ \; + \; N_-}.
\end{eqnarray}
%
%
This asymmetry does not require the prompt lepton assignment needed for calculating the usual $A_{FB}$ for the underlying production process $pp\rightarrow l^{+} N$, also avoiding the identical colliding beams problem \cite{delAguila:2008ir}. As can be seen in Fig.\ref{fig:asidv5n}, where we show the angular distribution for a mass of $m_N=5 ~GeV$, the contributions from both the scalar and vector operators lead to an imbalance. However, we expect the angular distributions to be related to the Dirac-Lorentz structure of the effective operators. Thus we expect a higher angular dependence in the vectorial case, which is visible in Fig.\ref{fig:asidv5n}.

\begin{figure*}
\centering
\subfloat[Contribution of the Vectorial Operators.]{\label{fig:asiV}\includegraphics[width=0.5\textwidth]{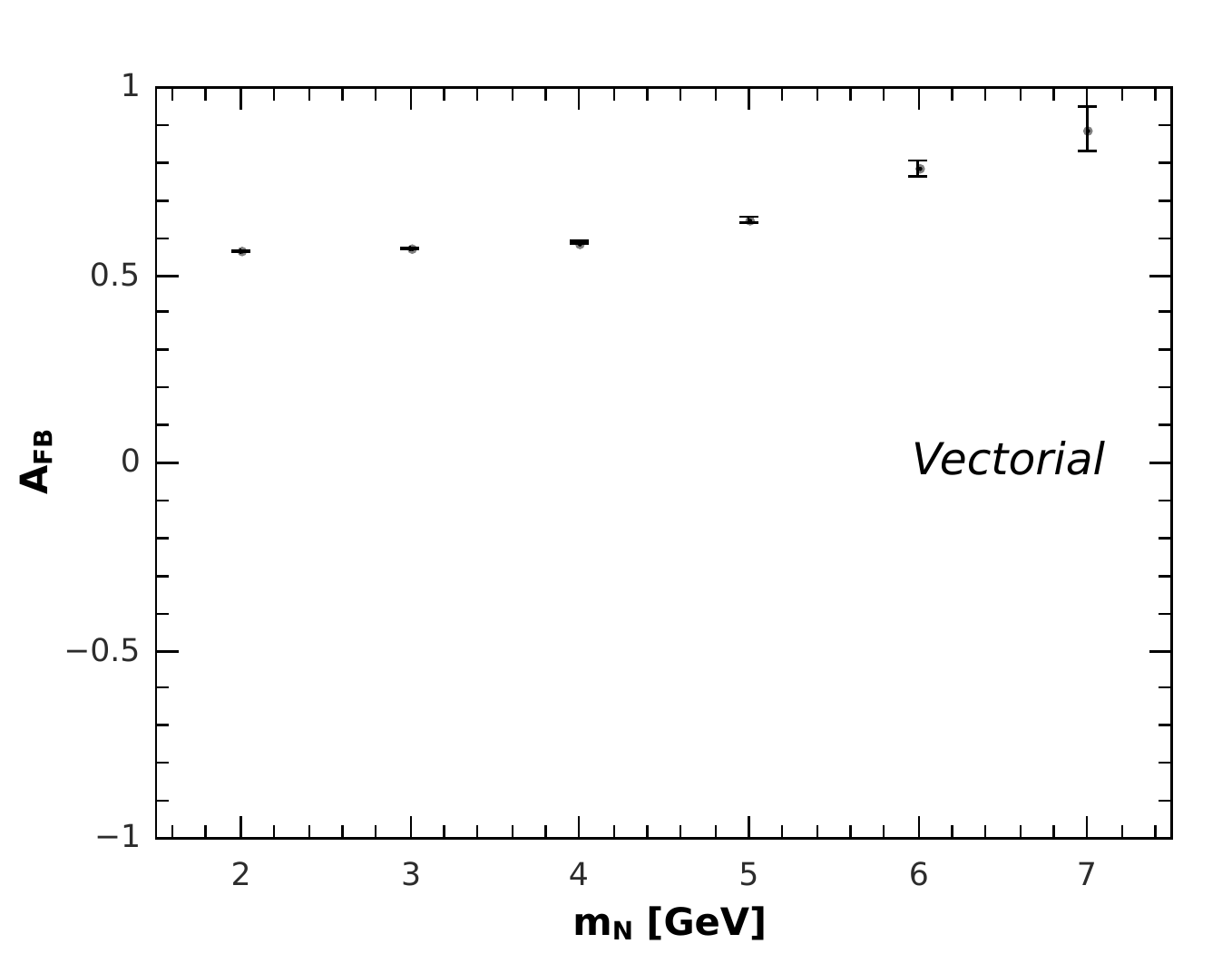}}
\subfloat[Contribution of the Scalar Operators.]{\label{fig:asiS}\includegraphics[width=0.5\textwidth]{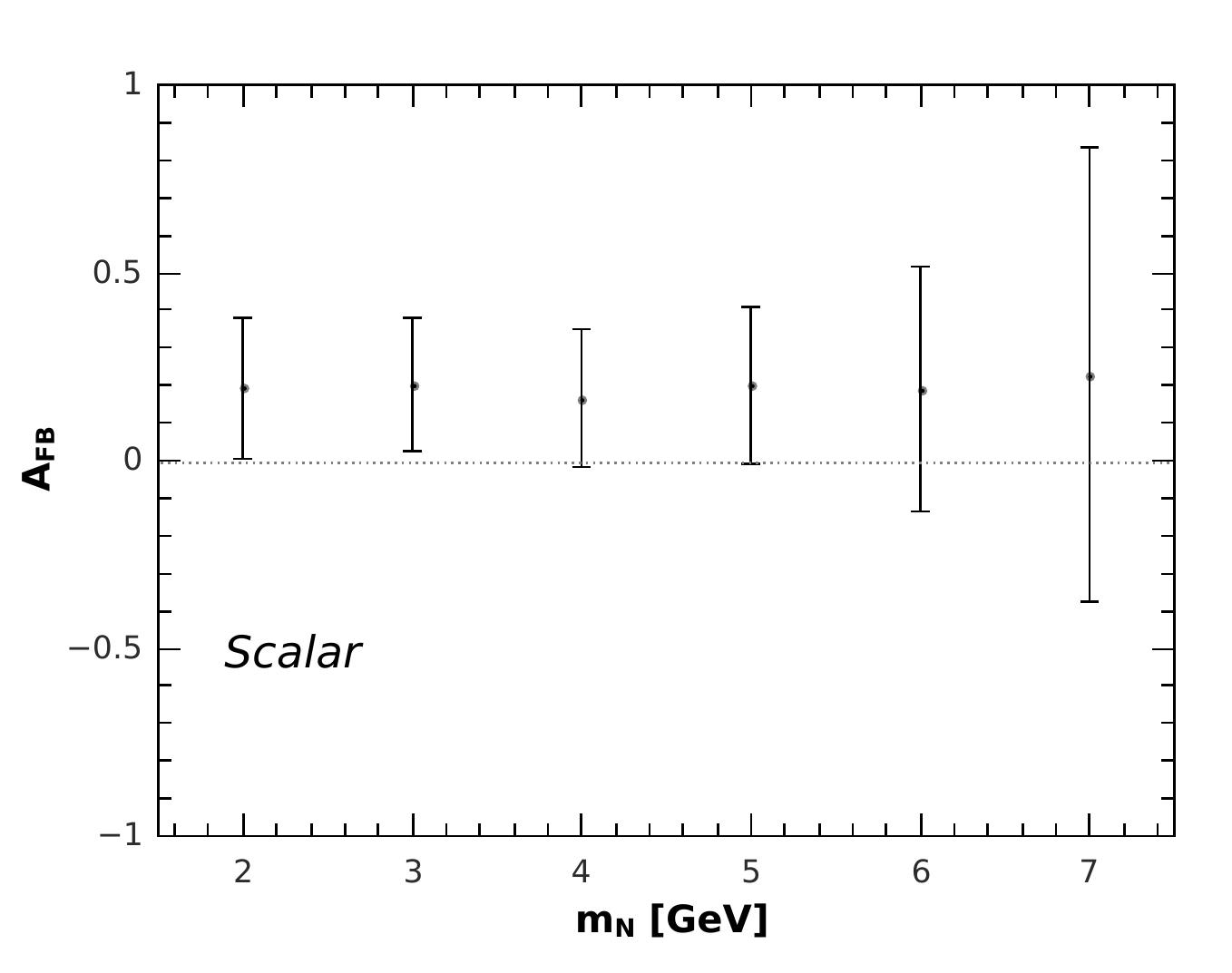}}
 \caption{\label{fig:asi} Asymmetry in the angular distribution between the final leptons ($l_1^{+} l_2^{+}= e\mu$, $\mu e$, $e e$, $\mu \mu$), with the errors estimates as defined in the text, for an integrated luminosity of $100 fb^{-1}$ and effective couplings in the set 0 ({\bf s0}).}
\end{figure*}

In order to estimate the chances of disentangling the contributions corresponding to the scalar and vector effective operators, we study the angular asymmetry $A^{l^+l^+}_{FB}$, taking into account the error 
\begin{eqnarray}
\Delta A^{l^+l^+}_{FB}=\sqrt{\left(\frac{\partial A^{l^+l^+}_{FB}}{\partial N_+}\right)^2 \left( \Delta N_+ \right)^2
\, + \, \left(\frac{\partial A^{l^+l^+}_{FB}}{\partial N_-}\right)^2 \left(\Delta N_- \right)^2}.
\end{eqnarray}
Assuming the number of events to be Poisson distributed, we write
\begin{equation}
\Delta N_+=\sqrt{N_+} \;\; \mbox{and} \;\; \Delta N_-=\sqrt{N_-}
\end{equation}
and a straightforward calculation leads to
\begin{eqnarray}
\Delta A^{l^+l^+}_{FB} = \sqrt{ \frac{1-(A^{l^+l^+}_{FB})^2}{N_++N_-}}.
\end{eqnarray}
The results for the $A^{l^+l^+}_{FB}$ observable for the vectorial and scalar operators cases are shown in Figs. \ref{fig:asiV} and \ref{fig:asiS}. Here we considered an integrated luminosity of $100 fb^{-1}$.

We find that for the scalar set of operators the asymmetry is compatible with zero, but the vectorial set shows a clear effect, different from zero, by several standard deviations. In the case of the scalar operators contribution, this is due to the bigger error bars, which in turn stem from the fact that the scalar operators give a lower contribution to the signal's cross section.     

\section{Prompt lepton and displaced photon  \\ plus missing $E_{T}$ signal}\label{sect:photon_neutrino}

As we found in \cite{Duarte:2015iba}, the $N\rightarrow \nu \gamma$ decay channel dominates the low $m_N$ region and is driven by one-loop generated tensorial operators. If one includes these in the $N$ decay width calculation, this decay mode would overshadow the ss-dilepton signal. So in the case that new physics involving tensorial modes is present, in this section we explore the chances to observe the $pp\rightarrow l^{+} \nu \gamma$ process at the LHC.  

The occurrence of magnetic moments giving $N-\nu$ transitions in the effective approach was studied in detail in \cite{Aparici:2009fh}. In that early work, the authors introduce three right handed Majorana neutrinos -one for each generation- and thus have a nonzero dimension 5 operator contribution $\mathcal{O}^{(5)}_{NB}=(\bar{N}~\zeta~\sigma^{\mu\nu}N^{c})B_{\mu\nu}$, due to the introduction of an antisymmetric matrix $\zeta$ in flavor space. The relevance of the $N\rightarrow \nu \gamma$ decay mode was discussed there, and the possible identification of the heavy neutrinos through the presence of displaced photons was hypothesized. Astrophysical bounds stemming from the cooling of red giant stars and supernovas on the coupling $\zeta$ were found for $m_N \ll 30~MeV$\cite{Aparici:2009fh}, well below the mass range considered here.

In our current approach, the magnetic moment is generated by the tensorial dimension 6 one-loop level generated operators in \eqref{eq:Ope-1-loop}. It leads to a very interesting phenomenology, not yet studied in the context of the LHC. In this section we perform a rapid calculation in order to grasp the possibility of detecting the sterile $N$ through the use of displaced photons observables, and explore the possibility of introducing cuts on the vertex displacement to enhance the signal observation.    
\begin{figure*}[h!]
\centering
\includegraphics[width=0.7\textwidth]{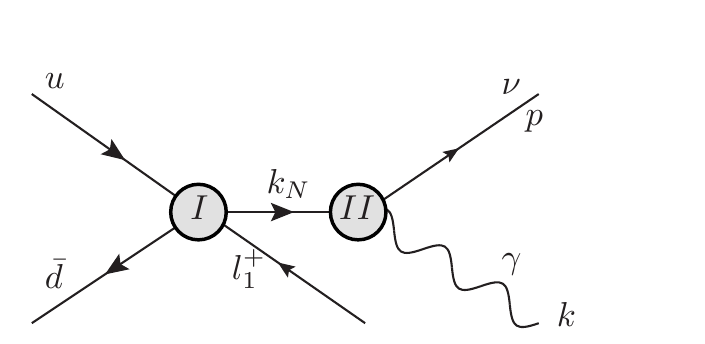}
 \caption{\label{fig:uD_Nl_lnug} Schematic momenta in eqs.\eqref{MI} and \eqref{MII}.}
\end{figure*}

The cross section for the production and decay of the heavy Majorana neutrino hard process we study can be written as:
\begin{eqnarray*}
 \hat{\sigma}_{u \bar{d}\rightarrow l^{+} \nu \gamma}(\hat{s})= \int \frac{1}{2\hat{s}} \overline{|\mathcal{M}|}^2 d\Phi_{3}.
\end{eqnarray*}
Here the squared scattering amplitude in the narrow width approximation is \cite{Uhlemann:2008pm}
\begin{eqnarray*}
 \overline{|\mathcal{M}|}^2= \frac{\pi}{2 \Gamma_{N}m_{N}} \delta(k_{N}^2-m_N^2)|\mathcal{M}^I|^2 |\mathcal{M}^{II}|^2  
\end{eqnarray*}
with the amplitude for the production vertex $I$ in Fig.\ref{fig:uD_Nl_lnug} being
\begin{eqnarray}\label{MI}
 |\mathcal{M}^I|^2&=&\frac{4}{\Lambda^4}\left[ (l_1.k_{N})(p_{d}.p_{u})[\alpha_{S_1}^{(i)2}+ \alpha_{S_2}^{(i)}(\alpha_{S_2}^{(i)}-\alpha_{S_3}^{(i)})]\right.
\nonumber
\\ && \left. + (l_1.p_{u})(p_{d}.k_{N})[\alpha_{S_2}^{(i)}\alpha_{S_3}^{(i)}+4 \alpha_{V_{0}}^{(i)2}] \right.
\nonumber
\\ && \left. +(l_1.p_{d})(p_{u}.k_{N})[\alpha_{S_3}^{(i)}(\alpha_{S_3}^{(i)}-\alpha_{S_2}^{(i)})+4 \alpha_{W}^{(i)2} |\Pi_{W}|^2]\right]
\end{eqnarray}
($i$ is the family index for the final charged anti-lepton $l_1^{+}$) and the amplitude for the decay vertex $II$ being
\begin{eqnarray}\label{MII}
 |\mathcal{M}^{II}|^2=\frac{16 v^2}{\Lambda^4}\left[(k.p)^2(\alpha_{NB}^{(j)} c_{W}+\alpha_{NW}^{(j)} s_{W})^{2}\right]
\end{eqnarray}
where $j=1,2$ is the family index for the final neutrino.

Again, the full cross section $\sigma(pp\rightarrow l^{+} \nu \gamma)$ is calculated making a Monte Carlo simulation, using the the CTEQ pdf functions \cite{Pumplin:2002vw}.

\subsubsection{Backgrounds}

The backgrounds for the $l^{+} \nu \gamma$ final state have been studied in SUSY searches in events with a photon, a lepton and missing transverse momentum at the LHC \cite{CMS:2015loa} for high momentum regions and considering signals where all the final particles are produced promptly. It is classified as misidentified photons or leptons and SM-electroweak background. The first arises when electrons or jets are misidentified as photons (coming from events where a photon decays into an $e^{+}e^{-}$ pair and an electron fails to register track seeds due to detector inefficiencies) or a jet is misidentified as a photon when a large fraction of its energy is carried by mesons decaying into photons. Misidentified leptons are reconstructed leptons not arising from $W$ or $Z$ boson decays, and come primarily from heavy-flavor quark decays and hadrons misidentified as leptons.

The SM-electroweak background is dominated by $Z,W -\gamma$ events that have the same signature as the signal events: a photon, a lepton and missing transverse energy ($E^{miss}_{T}$) from neutrinos. In the case of the dominant SM-electroweak background, leptons arise mostly from $W\rightarrow \nu l$ decays, and the $E^{miss}_{T}$ peaks nearly at $m_{W}/2$. In our case, for few-$GeV$ $m_N$ this background could be reduced imposing a cut on the $\nu-\gamma$ system transverse mass $M_{T}(\nu, \gamma) \lesssim m_N$ as defined in \cite{CMS:2015loa}, but changing the role of the charged lepton with the photon (see \cite{Khachatryan:2016hns}), given they come from the decaying $N$.

\begin{figure*}[h!]
\centering
\includegraphics[width=0.5\textwidth]{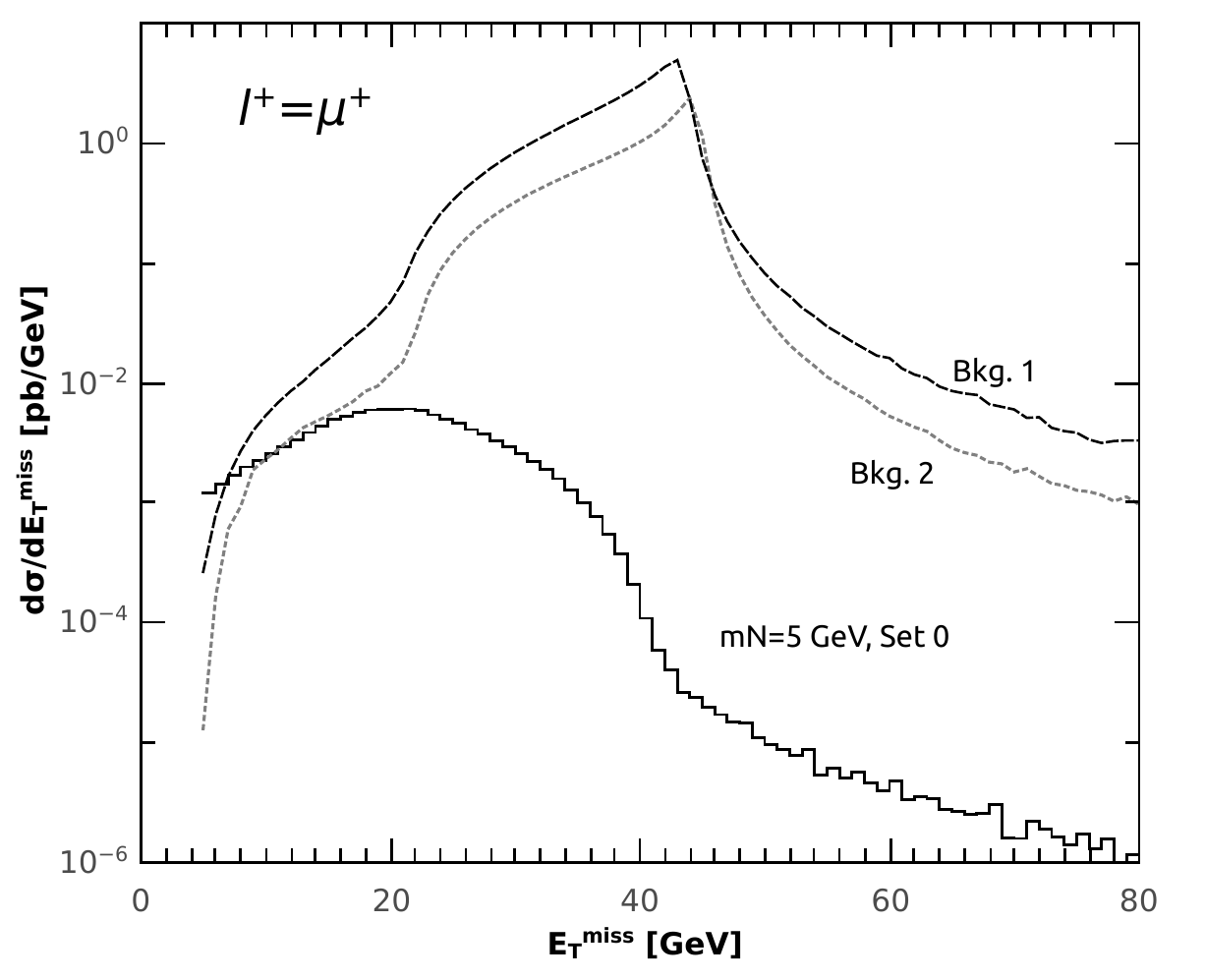}
 \caption{\label{fig:mpT_bkg_dist} Missing transverse energy $E^{miss}_{T}$ distribution for the dominant SM-electroweak background, and the signal for $m_N=5 ~GeV$ and set 0 ({\bf{s0}}) in the muon channel.}
\end{figure*}
We made a fast parton level simulation of the dominant SM-electroweak background with the CalcHep code \cite{Belyaev:2012qa}. As an illustration, in Fig. \ref{fig:mpT_bkg_dist} we show the $E^{miss}_{T}$ distribution curves for the dominant SM-electroweak background and the signal for $m_N=5 ~GeV$ and couplings set 0 ({\bf{s0}}) in the $l^{+}=\mu^{+}$ channel. The curve labeled $Bkg.1$ shows the background including only a kinematical cut $|\eta|\leq 1.47$ for the final particles, and a separation between the prompt lepton and the photon in the lab frame: $|cos(\theta_{\ell \gamma})|\leq 0.9$. The curve labeled $Bkg.2$ includes also a cut on the $\nu-\gamma$ system transverse mass $M_{T}(\nu, \gamma) \leq m_N= 5~GeV$. Here we do not consider the acceptance factor $P_{DI}$ in \eqref{eq:Nevents}. 

We find the signal $E^{miss}_{T}$ distribution is not very sensitive to the $m_N$ changes in the few-$GeV$ $m_N$ region we study. This would allow to also impose a cut $E^{miss}_{T}\gtrsim m_N$ to reduce possible backgrounds without real $E^{miss}_{T}$. The fake $E^{miss}_{T}$ backgrounds suppression depends mainly on the resolution in the $E^{miss}_{T}$ soft-terms, which are sensitive to the effects of pile-up in the $E^{miss}_{T}$ measurements. Its value is around $5-10 ~GeV$ \cite{ATL-PHYS-PUB-2015-027, Aad:2016nrq, Khachatryan:2014gga}.

In Table \ref{tab:ETmisscut} we show the number of signal events for $m_N=5 ~GeV$ and for the $Bkg.2$ when imposing different $E^{miss}_{T}$ cuts, for $\mathcal{L}=100~fb^{-1}$. The signal number of events is over the background in the optimistic interval $5~GeV \leq E^{miss}_{T} \leq 10 ~GeV$, which leaves a small phase space window for observations.

We have not quantitatively studied the fake $E^{miss}_{T}$ backgrounds suppression using a minimum $E^{miss}_{T}$ cut. Instead, our proposal is to study the possibilities to suppress the backgrounds imposing cuts that take into account the vertex displacement of the signal, as will be shown in the next sections.

\begin{table}[t]
 \centering
 \begin{tabular}{l c c r}
\firsthline
$E^{miss}_{T}$ cut & Signal events ($m_N=5 ~GeV$)~ ~& $Bkg.2$ events ~ ~&  Sig.$/$ Bkg. ($\%$) \\
\hline
 $5~GeV \leq E^{miss}_{T} \leq 10 ~GeV$ & $1\times 10^3$  & $6\times 10^2$	 & $167$ \\
 $5~GeV \leq E^{miss}_{T} \leq 20 ~GeV$ & $6\times 10^3$  & $7\times 10^3$ 	 & $86$ \\
 $5~GeV \leq E^{miss}_{T} \leq 30 ~GeV$ & $1\times 10^4$  & $1.5 \times 10^5$    & $7$ \\
\lasthline
 \end{tabular}
\caption{$E^{miss}_{T}$ cut for $l^{+}=\mu^{+}$, set 0 ({\bf{s0}}) and $\mathcal{L}=100~fb^{-1}$. }\label{tab:ETmisscut}
\end{table}

\subsubsection{Signal}

For our signal, the smoking gun for the production and flight of the $N$ is its decay length, that leads to a finite separation between the primary and secondary vertex. We are mostly interested in exploiting this distinctive displaced vertex feature, in this case measuring the distance between the prompt lepton track and the displaced photon flight direction, in order to reduce the backgrounds in the low $m_N$ region.

\begin{figure*}[h!]
\centering
\subfloat{\includegraphics[width=0.45\textwidth]{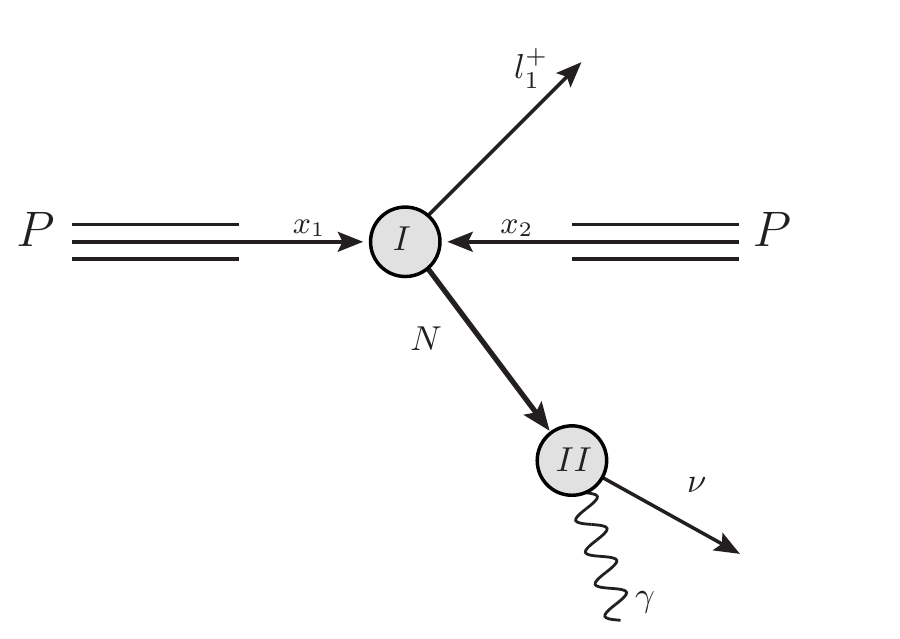}}
~
\subfloat{\includegraphics[width=0.45\textwidth]{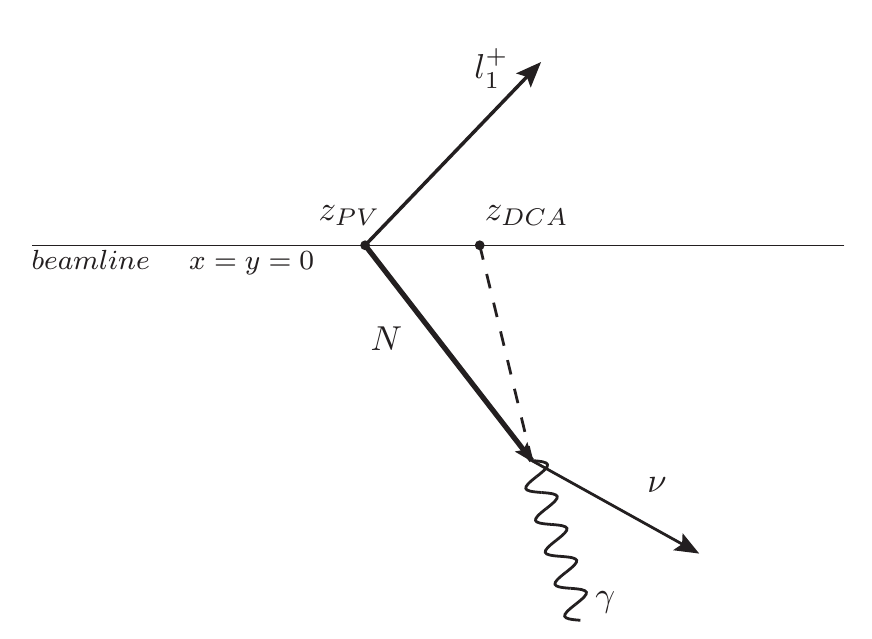}}
 \caption{\label{z_sketch} Schematic representation of the $pp\rightarrow l^{+} \nu \gamma$ process and the $z_{DCA}$ observable.}
\end{figure*}

In the LHC, observables using non-pointing photons have been defined and new physics searches were performed involving photons originating from a displaced vertex due to the decay of a long-lived particle into a photon and an invisible particle. These non-pointing photons and $E^{miss}_{T}$ final state searches \cite{Aad:2013oua, Aad:2014gfa} focus on the discovery of long-lived SUSY particles, but as is the case for this work, could serve the purpose of discovering heavy Majorana neutrinos as well. The analysis technique developed exploits the capabilities of the ATLAS electromagnetic calorimeter to make precise measurements of the flight direction of photons. This direction can be determined by measuring precisely the lateral and the longitudinal positions of the photon-originated shower in the front and middle layers of the EM calorimeter. The variable used as a measure of the degree of non-pointing of the photon is called $z_{DCA}$, the difference between the $z$ coordinate of the photon extrapolated back to its distance of closest approach ($DCA$) to the beamline (i.e. $x=y=0$) and $z_{PV}$, the $z$ coordinate of the primary vertex \cite{Nikiforou:2014cka}, as shown in Fig. \ref{z_sketch}.

\subsubsection{Numerical results}

The distance of closest approach $z_{DCA}$ of the displaced photon can be obtained in our Monte Carlo simulation using the simulated momenta of the outgoing photons and the decay length of the Majorana neutrino defined in \eqref{eq:L_N}, which is obtained from the simulated momenta of the Majorana neutrinos, and the exponentially sampled distance $\ell$. 

\begin{figure*}[h!]
\centering
\subfloat{\includegraphics[width=0.49\textwidth]{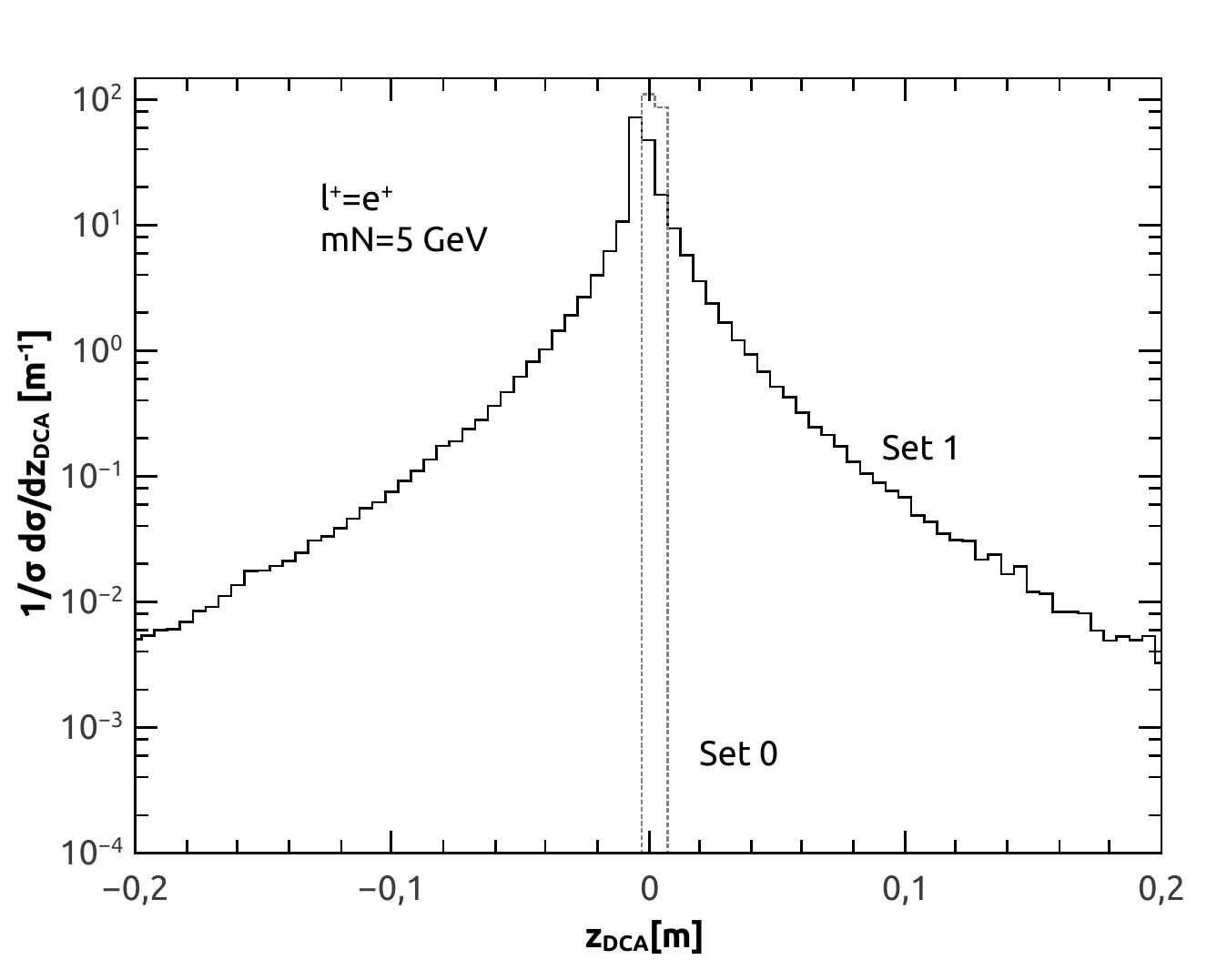}}
~
\subfloat{\includegraphics[width=0.49\textwidth]{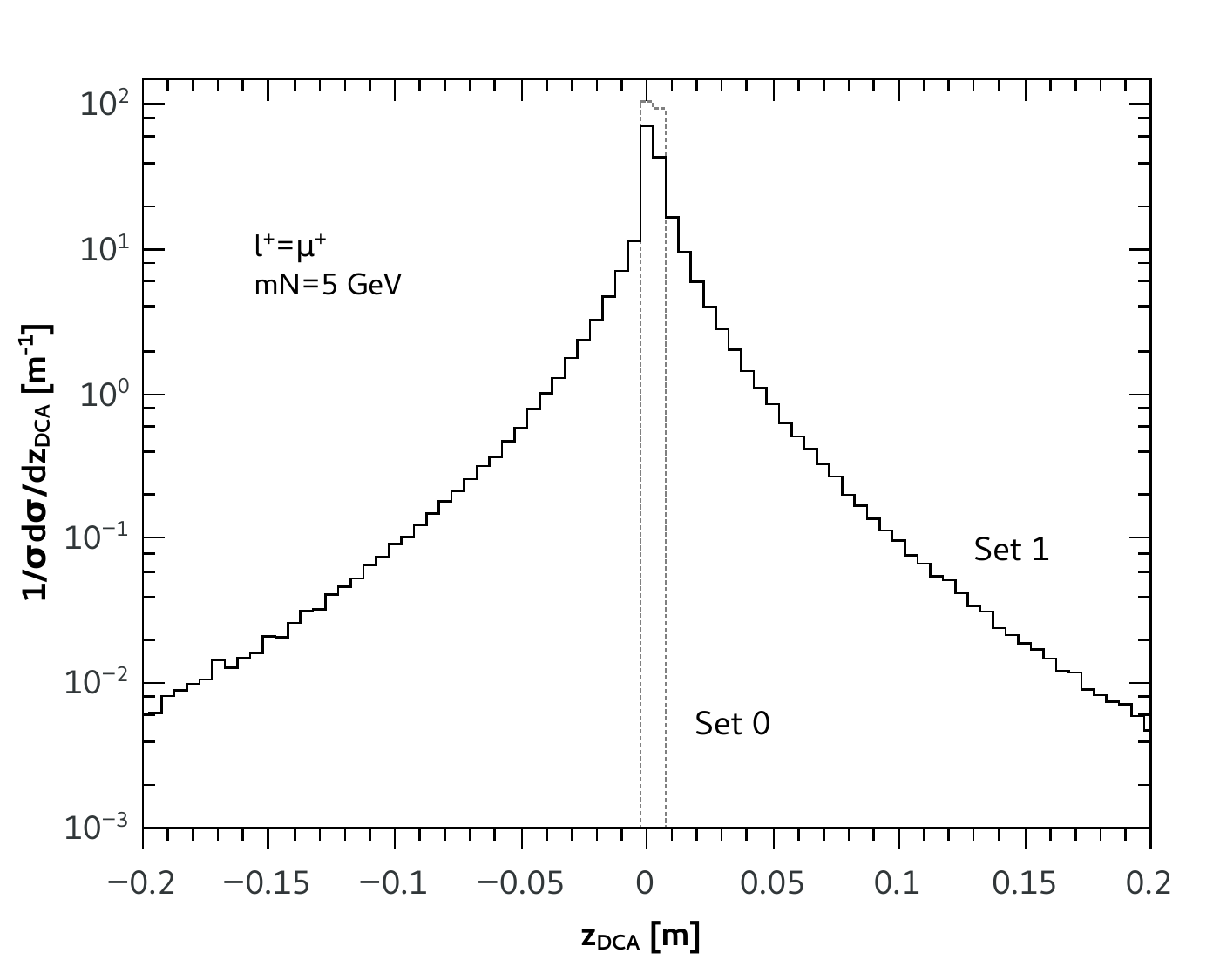}}

\subfloat{\includegraphics[width=0.49\textwidth]{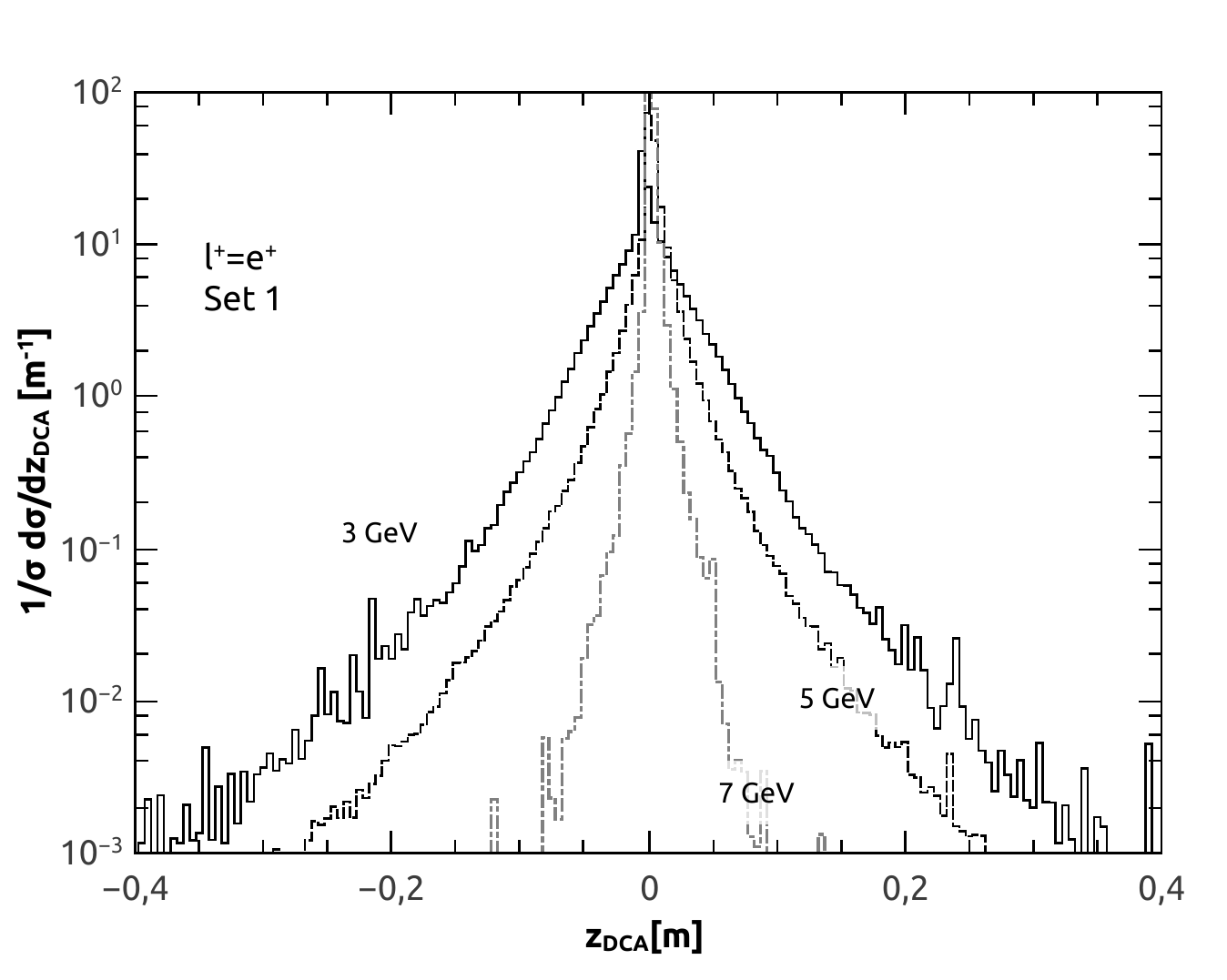}}
~
\subfloat{\includegraphics[width=0.49\textwidth]{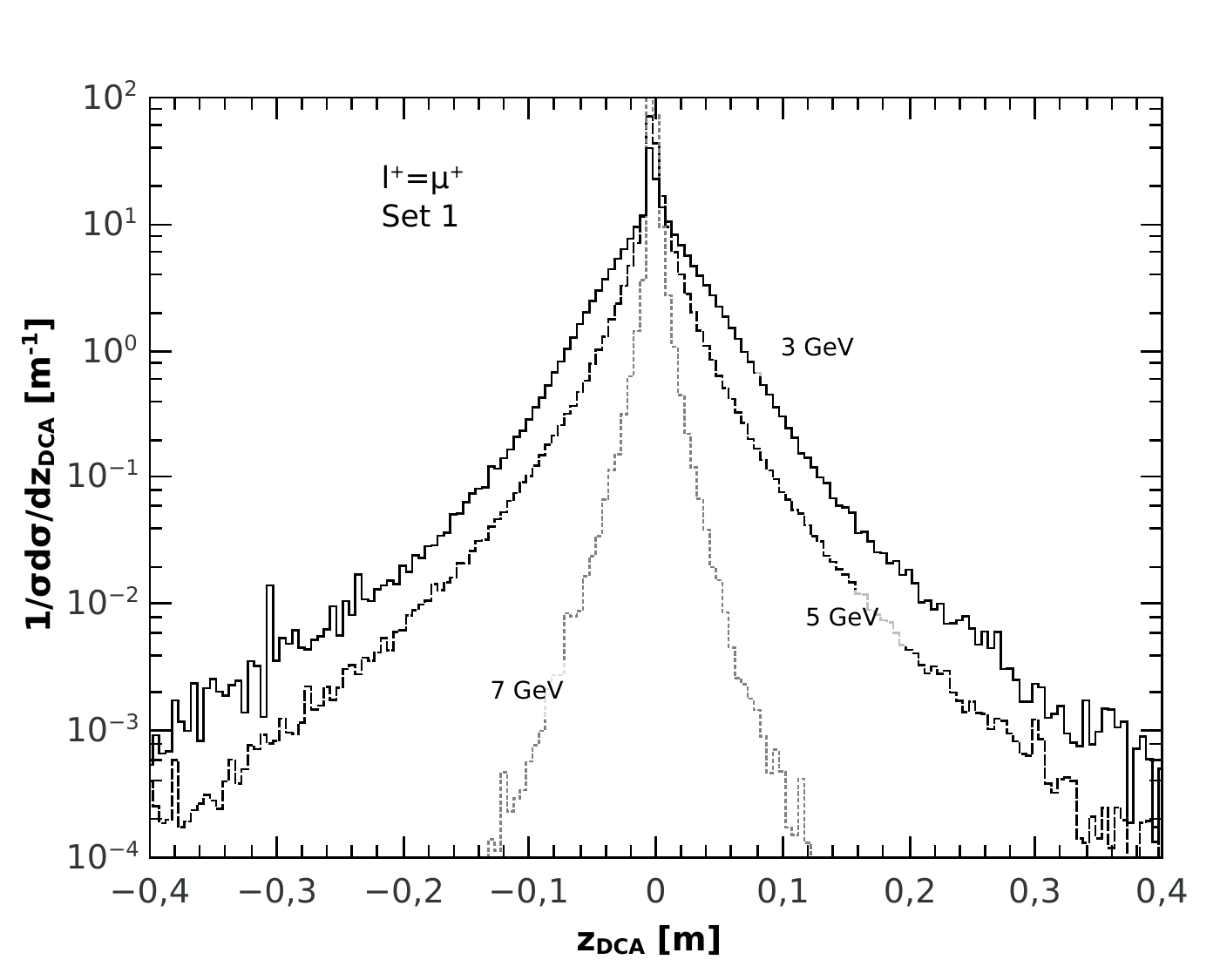}}
 \caption{\label{fig:z_dca} $z_{DCA}$ distribution for the $l^{+} \nu \gamma$ channel, for distinct $m_N$ values and effective couplings sets 0 ({\bf s0}) and 1 ({\bf s1}).}
\end{figure*}

Taking $z_{PV}=0$, in Fig.\ref{fig:z_dca} we plot the normalized differential cross section of the signal as a function of the distance of closest approach $z_{DCA}$  for the $l^{+}= e^{+},\mu^{+}$ channels. Here we have introduced only one kinematical cut, taking $|\eta|\leq 1.47$ for the final particles, and a separation between the prompt lepton and the photon in the lab frame: $|cos(\theta_{\ell \gamma})|\leq 0.9$. In the lower panels of Fig.\ref{fig:z_dca} we show the $z_{DCA}$ distribution for different mass values $m_N$ and the couplings set 1 ({\bf{s1}}). It can be seen from this pictures that the $z_{DCA}$ distribution is very sensitive to both the couplings set and the Majorana neutrino masses. In this plots we also consider the probability for the $N$ decay to occur inside the detector $P_{DI}$ defined in Sect.\ref{sect:low_mN} with a detector size $L_{D}=1m$.  

%

%

\begin{figure*}[h!]
\centering
{\includegraphics[width=\textwidth]{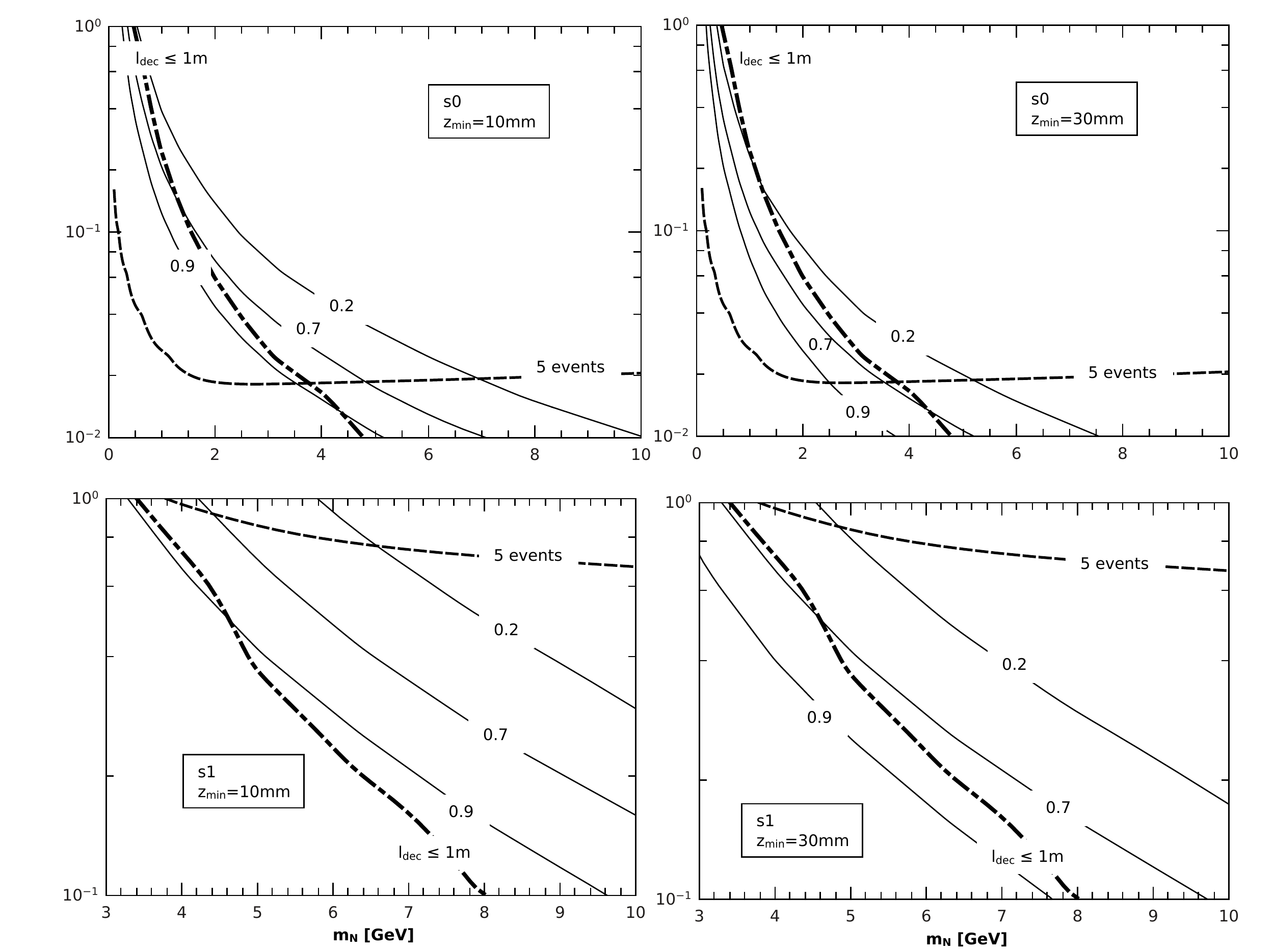}}
 \caption{\label{fig:z_dca_reg} Contour levels of the quotient $\mathcal{Q}$ defined in \eqref{eq:Q} in the $(\alpha,m_N)$ plane for the $z_{DCA}$ distribution for the $\mu^{+} \nu \gamma$ channel. See the text for reference.}
\end{figure*}

In order to show more explicitly the dependence of the $z_{DCA}$ distribution on the values of the effective couplings for distinct masses, and the efficiency of a cut on a minimal distance cut $z_{min}$, we consider the level contours of the quotient $\mathcal{Q}$ in the $(\alpha,m_N)$ plane in Fig.\ref{fig:z_dca_reg}
\begin{eqnarray}\label{eq:Q}
 \mathcal{Q}=\frac{\int_{z_{min}}^{z_{D}} \frac{d\sigma}{dz}  ~dz}{\int_{0}^{z_{D}} \frac{d\sigma}{dz}  ~dz}.
\end{eqnarray}
Here in the integrands $z$ stands for $z_{DCA}$, and in the numerator we consider the number of events with $z_{DCA}$ between a minimal distance cut $z_{min}$ and a fiducial detector size, which for simplicity we take as $z_D=1~m$. In the denominator we consider all the events between $0$ and $z_D$. The integrated luminosity is canceled in the quotient.

When the distribution is very concentrated around $z_{DCA}=0$, discarding the events with $z_{DCA}\leq z_{min}$ will reduce the value of $\mathcal{Q}$. The level contours of this quotient tell us how the spreading depends on the values of the effective couplings $\alpha$ and the mass $m_N$, giving a measure of the signal efficiency for this cut. In Fig.\ref{fig:z_dca_reg} the vertical axes show the factor by which we multiply the values of the effective couplings defined for each coupling set in Tab.\ref{tab:alpha-sets}. A factor $1$ implies we consider the values of the experimental bounds discussed in Sec.\ref{sect:bounds}. The full lines show the contour levels $\mathcal{Q}=0.2, ~0.7, ~0.9$. We also plot the curve given by a decay length \footnote{Here, for simplicity, we take the Majorana neutrino decay length to be the average decay length $\ell_N$ in \eqref{eq:l_N}. } of the Majorana neutrinos equal to the detector size $\ell_{dec}=L_D= 1~m$ (dot-dashed line), and the line showing the values of couplings and masses leading to $5$ events (dashed line) considering an integrated luminosity $\mathcal{L}= 100 fb^{-1}$. 
The region of interest is the one over the $\ell_{dec}=L_D$ line (because higher coupling values lead to a lower decay length for the Majorana $N$) and also over the $5-$events curves (because higher couplings give more events). The first constraint ensures the Majorana neutrino decays inside the detector volume. In each plot the couplings set and the value of the $z_{min}$ cut is indicated. We show the curves for $z_{min}=10 ~mm$ and $z_{min}=30 ~mm$ \footnote{The expected pointing resolution for $|z_{DCA}|< 100 ~mm$ is in the $10-20 ~mm$ range \cite{Aad:2013oua}.}. We find regions with events where the signal has high efficiency after the $z_{min}$ cut.

\begin{figure*}[h!]
\centering
\subfloat{\includegraphics[width=0.45\textwidth]{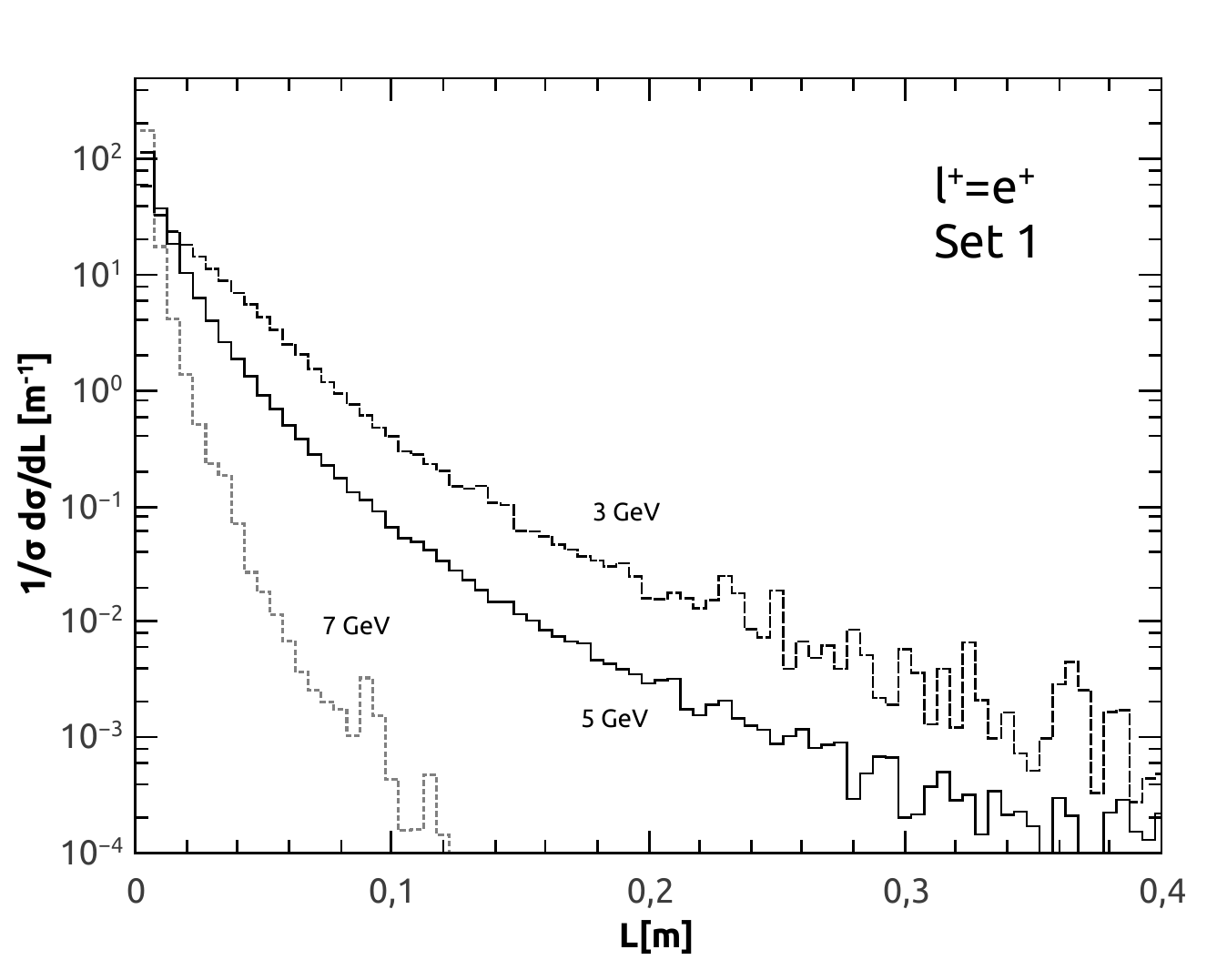}}
~
\subfloat{\includegraphics[width=0.45\textwidth]{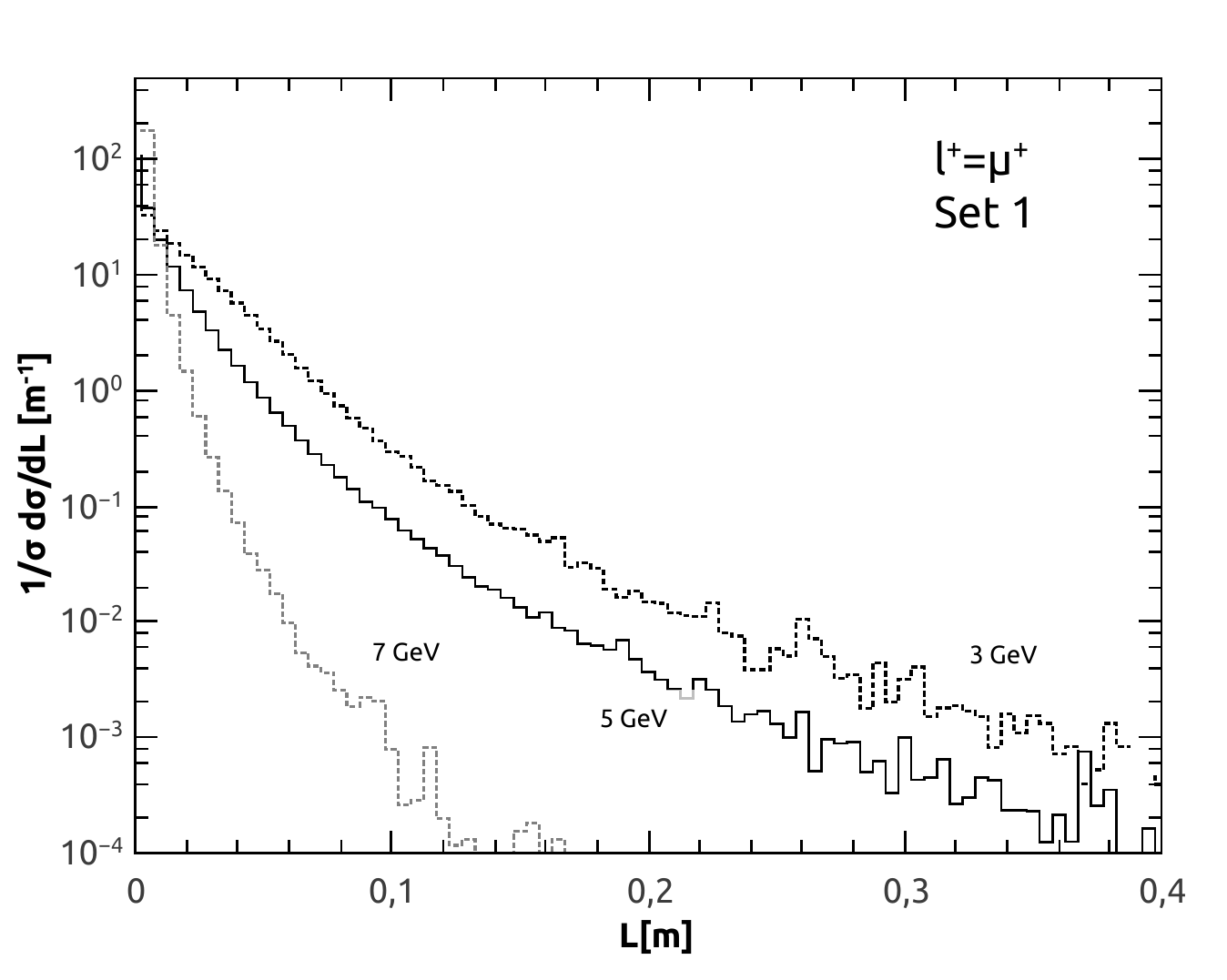}}
 \caption{\label{fig:ddv} $L^{l^{+}\gamma}$ differential distribution for the $l^{+} \nu \gamma$ channel.}
\end{figure*}

We also study the distribution of the signal with the distance $L^{l^{+}\gamma}$ between the prompt lepton and the displaced photon traces defined equivalently as in Sec.\ref{sect:ss_dilepton}. In Fig.\ref{fig:ddv} we show the normalized differential cross section of the signal for the $l^{+}=e^{+},\mu^{+}$ channels as a function of the displacement distance $L^{l^{+}\gamma}$. Here we also take into account the probability $P_{DI}$. 

We study the efficiency of the signal after a cut $L^{l^{+}\gamma}>5 ~mm$, for different values of $m_N$ and the couplings set 1 ({\bf s1}), considering an integrated luminosity $\mathcal{L}=100 fb^{-1}$, for the $l^{+}= \mu^{+}$ channel. Our results are shown in Tab.\ref{eff_ddv}.
\begin{table}[t]
 \centering
 \begin{tabular}{c r }
\firsthline
$m_N$ & Sig.eff. ($L^{l^{+}\gamma}$ cut) \\
\hline
 $3$ $GeV$ & $66.1$ $\%$ \\
 $5$ $GeV$ & $76.7$ $\%$ \\
 $7$ $GeV$ & $82.0$ $\%$ \\
\lasthline
 \end{tabular}
\caption{Signal efficiency after cut $L^{l^{+}\gamma}>5 ~mm$ for the couplings set 1 ({\bf{s1}}).}\label{eff_ddv}
\end{table}
%


Finally, we would like to stress that information on the photon's time of flight used in LHC searches \cite{Aad:2014gfa, Aad:2013oua} could also be used to further reduce the backgrounds, relating it to the $\nu-\gamma$ system transverse mass.

\section{Final remarks}\label{sect:final}

The heavy neutrino effective field theory, parameterizing new high-scale weakly coupled physics beyond the minimal seesaw mechanism gives a model independent framework for studying possible lepton number violating effects in hadron colliders. In the simplified scenario with only one extra heavy SM singlet $N$ with effective interactions, it also leads to dimension 6 Lagrangian terms giving a magnetic moment for the heavy $N$. 

In this paper we study the effects of a heavy sterile Majorana neutrino with effective interactions in the  $pp\rightarrow l_i^{+}l_j^{+} jj$ and $pp\rightarrow l_i^{+} \nu \gamma$ processes in the LHC. For low $m_N$ masses of a few-$GeV$ the $N$ behaves as a long lived neutral particle, leading to a distinctive displaced vertex signature that can be used to reduce backgrounds at the LHC. 

Exploiting this fact we design a forward-backward like asymmetry for the  $pp\rightarrow l_i^{+}l_j^{+} jj$ channel, and find it could disentangle the contribution from scalar and vectorial effective operators, giving a hint on the kind of new physics contributing to the effective interactions. Making a statistical analysis we find the contributions from the two operator groups could be distinguished. In particular, the vectorial operators give an asymmetry clearly distinct from zero.  

The heavy neutrino effective magnetic moment -yet unstudied in the LHC- leads in this mass range to a novel displaced photon signature in the $pp\rightarrow l_i^{+} \nu \gamma$ channel, which we find could be detected with the aid of the non-pointing photon observable $z_{DCA}$ with cuts in the displacement between the prompt lepton and the outgoing photon to reject the backgrounds.

Our encouraging findings suggest to deepen this preliminary parton level studies in order to accurately model the signals and backgrounds in the LHC with Monte Carlo simulations including hadronization and detector simulation techniques. The not-that-heavy Majorana neutrinos could be discovered at the LHC in the case the triggers and analysis for Run II dedicated searches keep reconstruction thresholds sufficiently low to efficiently tag displaced vertices from signal processes.   \\

{\bf Acknowledgments}

We thank CONICET (Argentina) and Universidad Nacional de Mar del
Plata (Argentina); and PEDECIBA, ANII, and CSIC-UdelaR (Uruguay) for their 
financial supports.

\bibliography{Bib_N_10_2017}

\begin{thebibliography}{92}
\expandafter\ifx\csname natexlab\endcsname\relax\def\natexlab#1{#1}\fi
\expandafter\ifx\csname bibnamefont\endcsname\relax
  \def\bibnamefont#1{#1}\fi
\expandafter\ifx\csname bibfnamefont\endcsname\relax
  \def\bibfnamefont#1{#1}\fi
\expandafter\ifx\csname citenamefont\endcsname\relax
  \def\citenamefont#1{#1}\fi
\expandafter\ifx\csname url\endcsname\relax
  \def\url#1{\texttt{#1}}\fi
\expandafter\ifx\csname urlprefix\endcsname\relax\def\urlprefix{URL }\fi
\providecommand{\bibinfo}[2]{#2}
\providecommand{\eprint}[2][]{\url{#2}}

\bibitem[{\citenamefont{Minkowski}(1977)}]{Minkowski:1977sc}
\bibinfo{author}{\bibfnamefont{P.}~\bibnamefont{Minkowski}},
  \bibinfo{journal}{Phys.Lett.} \textbf{\bibinfo{volume}{B67}},
  \bibinfo{pages}{421} (\bibinfo{year}{1977}).

\bibitem[{\citenamefont{Mohapatra and Senjanovic}(1980)}]{Mohapatra:1979ia}
\bibinfo{author}{\bibfnamefont{R.~N.} \bibnamefont{Mohapatra}}
  \bibnamefont{and}
  \bibinfo{author}{\bibfnamefont{G.}~\bibnamefont{Senjanovic}},
  \bibinfo{journal}{Phys.Rev.Lett.} \textbf{\bibinfo{volume}{44}},
  \bibinfo{pages}{912} (\bibinfo{year}{1980}).

\bibitem[{\citenamefont{Yanagida}(1980)}]{Yanagida:1980xy}
\bibinfo{author}{\bibfnamefont{T.}~\bibnamefont{Yanagida}},
  \bibinfo{journal}{Prog.Theor.Phys.} \textbf{\bibinfo{volume}{64}},
  \bibinfo{pages}{1103} (\bibinfo{year}{1980}).

\bibitem[{\citenamefont{Gell-Mann et~al.}(1979)\citenamefont{Gell-Mann, Ramond,
  and Slansky}}]{GellMann:1980vs}
\bibinfo{author}{\bibfnamefont{M.}~\bibnamefont{Gell-Mann}},
  \bibinfo{author}{\bibfnamefont{P.}~\bibnamefont{Ramond}}, \bibnamefont{and}
  \bibinfo{author}{\bibfnamefont{R.}~\bibnamefont{Slansky}},
  \bibinfo{journal}{Conf.Proc.} \textbf{\bibinfo{volume}{C790927}},
  \bibinfo{pages}{315} (\bibinfo{year}{1979}), \eprint{1306.4669}.

\bibitem[{\citenamefont{Schechter and Valle}(1980)}]{Schechter:1980gr}
\bibinfo{author}{\bibfnamefont{J.}~\bibnamefont{Schechter}} \bibnamefont{and}
  \bibinfo{author}{\bibfnamefont{J.~W.~F.} \bibnamefont{Valle}},
  \bibinfo{journal}{Phys. Rev.} \textbf{\bibinfo{volume}{D22}},
  \bibinfo{pages}{2227} (\bibinfo{year}{1980}).

\bibitem[{\citenamefont{Kayser et~al.}(1989)\citenamefont{Kayser, Gibrat-Debu,
  and Perrier}}]{Kayser:1989iu}
\bibinfo{author}{\bibfnamefont{B.}~\bibnamefont{Kayser}},
  \bibinfo{author}{\bibfnamefont{F.}~\bibnamefont{Gibrat-Debu}},
  \bibnamefont{and} \bibinfo{author}{\bibfnamefont{F.}~\bibnamefont{Perrier}},
  \bibinfo{journal}{World Sci.Lect.Notes Phys.} \textbf{\bibinfo{volume}{25}},
  \bibinfo{pages}{1} (\bibinfo{year}{1989}).

\bibitem[{\citenamefont{Atre et~al.}(2009)\citenamefont{Atre, Han, Pascoli, and
  Zhang}}]{Atre:2009rg}
\bibinfo{author}{\bibfnamefont{A.}~\bibnamefont{Atre}},
  \bibinfo{author}{\bibfnamefont{T.}~\bibnamefont{Han}},
  \bibinfo{author}{\bibfnamefont{S.}~\bibnamefont{Pascoli}}, \bibnamefont{and}
  \bibinfo{author}{\bibfnamefont{B.}~\bibnamefont{Zhang}},
  \bibinfo{journal}{JHEP} \textbf{\bibinfo{volume}{0905}}, \bibinfo{pages}{030}
  (\bibinfo{year}{2009}), \eprint{0901.3589}.

\bibitem[{\citenamefont{del Aguila et~al.}(2007)\citenamefont{del Aguila,
  Aguilar-Saavedra, and Pittau}}]{delAguila:2007qnc}
\bibinfo{author}{\bibfnamefont{F.}~\bibnamefont{del Aguila}},
  \bibinfo{author}{\bibfnamefont{J.~A.} \bibnamefont{Aguilar-Saavedra}},
  \bibnamefont{and} \bibinfo{author}{\bibfnamefont{R.}~\bibnamefont{Pittau}},
  \bibinfo{journal}{JHEP} \textbf{\bibinfo{volume}{10}}, \bibinfo{pages}{047}
  (\bibinfo{year}{2007}), \eprint{hep-ph/0703261}.

\bibitem[{\citenamefont{del Aguila et~al.}(2009)\citenamefont{del Aguila,
  Bar-Shalom, Soni, and Wudka}}]{delAguila:2008ir}
\bibinfo{author}{\bibfnamefont{F.}~\bibnamefont{del Aguila}},
  \bibinfo{author}{\bibfnamefont{S.}~\bibnamefont{Bar-Shalom}},
  \bibinfo{author}{\bibfnamefont{A.}~\bibnamefont{Soni}}, \bibnamefont{and}
  \bibinfo{author}{\bibfnamefont{J.}~\bibnamefont{Wudka}},
  \bibinfo{journal}{Phys.Lett.} \textbf{\bibinfo{volume}{B670}},
  \bibinfo{pages}{399} (\bibinfo{year}{2009}), \eprint{0806.0876}.

\bibitem[{\citenamefont{Kovalenko et~al.}(2009)\citenamefont{Kovalenko, Lu, and
  Schmidt}}]{Kovalenko:2009td}
\bibinfo{author}{\bibfnamefont{S.}~\bibnamefont{Kovalenko}},
  \bibinfo{author}{\bibfnamefont{Z.}~\bibnamefont{Lu}}, \bibnamefont{and}
  \bibinfo{author}{\bibfnamefont{I.}~\bibnamefont{Schmidt}},
  \bibinfo{journal}{Phys. Rev.} \textbf{\bibinfo{volume}{D80}},
  \bibinfo{pages}{073014} (\bibinfo{year}{2009}), \eprint{0907.2533}.

\bibitem[{\citenamefont{Alva et~al.}(2015)\citenamefont{Alva, Han, and
  Ruiz}}]{Alva:2014gxa}
\bibinfo{author}{\bibfnamefont{D.}~\bibnamefont{Alva}},
  \bibinfo{author}{\bibfnamefont{T.}~\bibnamefont{Han}}, \bibnamefont{and}
  \bibinfo{author}{\bibfnamefont{R.}~\bibnamefont{Ruiz}},
  \bibinfo{journal}{JHEP} \textbf{\bibinfo{volume}{02}}, \bibinfo{pages}{072}
  (\bibinfo{year}{2015}), \eprint{1411.7305}.

\bibitem[{\citenamefont{Das and Okada}(2016)}]{Das:2015toa}
\bibinfo{author}{\bibfnamefont{A.}~\bibnamefont{Das}} \bibnamefont{and}
  \bibinfo{author}{\bibfnamefont{N.}~\bibnamefont{Okada}},
  \bibinfo{journal}{Phys. Rev.} \textbf{\bibinfo{volume}{D93}},
  \bibinfo{pages}{033003} (\bibinfo{year}{2016}), \eprint{1510.04790}.

\bibitem[{\citenamefont{Dib and Kim}(2015)}]{Dib:2015oka}
\bibinfo{author}{\bibfnamefont{C.~O.} \bibnamefont{Dib}} \bibnamefont{and}
  \bibinfo{author}{\bibfnamefont{C.~S.} \bibnamefont{Kim}},
  \bibinfo{journal}{Phys. Rev.} \textbf{\bibinfo{volume}{D92}},
  \bibinfo{pages}{093009} (\bibinfo{year}{2015}), \eprint{1509.05981}.

\bibitem[{\citenamefont{Izaguirre and Shuve}(2015)}]{Izaguirre:2015pga}
\bibinfo{author}{\bibfnamefont{E.}~\bibnamefont{Izaguirre}} \bibnamefont{and}
  \bibinfo{author}{\bibfnamefont{B.}~\bibnamefont{Shuve}},
  \bibinfo{journal}{Phys. Rev.} \textbf{\bibinfo{volume}{D91}},
  \bibinfo{pages}{093010} (\bibinfo{year}{2015}), \eprint{1504.02470}.

\bibitem[{\citenamefont{Chakdar et~al.}(2017)\citenamefont{Chakdar, Ghosh,
  Hoang, Hung, and Nandi}}]{Chakdar:2016adj}
\bibinfo{author}{\bibfnamefont{S.}~\bibnamefont{Chakdar}},
  \bibinfo{author}{\bibfnamefont{K.}~\bibnamefont{Ghosh}},
  \bibinfo{author}{\bibfnamefont{V.}~\bibnamefont{Hoang}},
  \bibinfo{author}{\bibfnamefont{P.~Q.} \bibnamefont{Hung}}, \bibnamefont{and}
  \bibinfo{author}{\bibfnamefont{S.}~\bibnamefont{Nandi}},
  \bibinfo{journal}{Phys. Rev.} \textbf{\bibinfo{volume}{D95}},
  \bibinfo{pages}{015014} (\bibinfo{year}{2017}), \eprint{1606.08502}.

\bibitem[{\citenamefont{Dube et~al.}(2017)\citenamefont{Dube, Gadkari, and
  Thalapillil}}]{Dube:2017jgo}
\bibinfo{author}{\bibfnamefont{S.}~\bibnamefont{Dube}},
  \bibinfo{author}{\bibfnamefont{D.}~\bibnamefont{Gadkari}}, \bibnamefont{and}
  \bibinfo{author}{\bibfnamefont{A.~M.} \bibnamefont{Thalapillil}},
  \bibinfo{journal}{Phys. Rev.} \textbf{\bibinfo{volume}{D96}},
  \bibinfo{pages}{055031} (\bibinfo{year}{2017}), \eprint{1707.00008}.

\bibitem[{\citenamefont{Das et~al.}(2016)\citenamefont{Das, Konar, and
  Majhi}}]{Das:2016hof}
\bibinfo{author}{\bibfnamefont{A.}~\bibnamefont{Das}},
  \bibinfo{author}{\bibfnamefont{P.}~\bibnamefont{Konar}}, \bibnamefont{and}
  \bibinfo{author}{\bibfnamefont{S.}~\bibnamefont{Majhi}},
  \bibinfo{journal}{JHEP} \textbf{\bibinfo{volume}{06}}, \bibinfo{pages}{019}
  (\bibinfo{year}{2016}), \eprint{1604.00608}.

\bibitem[{\citenamefont{Degrande et~al.}(2016)\citenamefont{Degrande,
  Mattelaer, Ruiz, and Turner}}]{Degrande:2016aje}
\bibinfo{author}{\bibfnamefont{C.}~\bibnamefont{Degrande}},
  \bibinfo{author}{\bibfnamefont{O.}~\bibnamefont{Mattelaer}},
  \bibinfo{author}{\bibfnamefont{R.}~\bibnamefont{Ruiz}}, \bibnamefont{and}
  \bibinfo{author}{\bibfnamefont{J.}~\bibnamefont{Turner}},
  \bibinfo{journal}{Phys. Rev.} \textbf{\bibinfo{volume}{D94}},
  \bibinfo{pages}{053002} (\bibinfo{year}{2016}), \eprint{1602.06957}.

\bibitem[{\citenamefont{Dev et~al.}(2014)\citenamefont{Dev, Pilaftsis, and
  Yang}}]{Dev:2013wba}
\bibinfo{author}{\bibfnamefont{P.~S.~B.} \bibnamefont{Dev}},
  \bibinfo{author}{\bibfnamefont{A.}~\bibnamefont{Pilaftsis}},
  \bibnamefont{and} \bibinfo{author}{\bibfnamefont{U.-k.} \bibnamefont{Yang}},
  \bibinfo{journal}{Phys. Rev. Lett.} \textbf{\bibinfo{volume}{112}},
  \bibinfo{pages}{081801} (\bibinfo{year}{2014}), \eprint{1308.2209}.

\bibitem[{\citenamefont{Ma and Pantaleone}(1989)}]{Ma:1989jpa}
\bibinfo{author}{\bibfnamefont{E.}~\bibnamefont{Ma}} \bibnamefont{and}
  \bibinfo{author}{\bibfnamefont{J.~T.} \bibnamefont{Pantaleone}},
  \bibinfo{journal}{Phys.Rev.} \textbf{\bibinfo{volume}{D40}},
  \bibinfo{pages}{2172} (\bibinfo{year}{1989}).

\bibitem[{\citenamefont{Buchmuller and Greub}(1991)}]{Buchmuller:1991tu}
\bibinfo{author}{\bibfnamefont{W.}~\bibnamefont{Buchmuller}} \bibnamefont{and}
  \bibinfo{author}{\bibfnamefont{C.}~\bibnamefont{Greub}},
  \bibinfo{journal}{Nucl.Phys.} \textbf{\bibinfo{volume}{B363}},
  \bibinfo{pages}{345} (\bibinfo{year}{1991}).

\bibitem[{\citenamefont{Hofer and Sehgal}(1996)}]{Hofer:1996cs}
\bibinfo{author}{\bibfnamefont{A.}~\bibnamefont{Hofer}} \bibnamefont{and}
  \bibinfo{author}{\bibfnamefont{L.}~\bibnamefont{Sehgal}},
  \bibinfo{journal}{Phys.Rev.} \textbf{\bibinfo{volume}{D54}},
  \bibinfo{pages}{1944} (\bibinfo{year}{1996}), \eprint{hep-ph/9603240}.

\bibitem[{\citenamefont{Peressutti et~al.}(2011)\citenamefont{Peressutti,
  Romero, and Sampayo}}]{Peressutti:2011kx}
\bibinfo{author}{\bibfnamefont{J.}~\bibnamefont{Peressutti}},
  \bibinfo{author}{\bibfnamefont{I.}~\bibnamefont{Romero}}, \bibnamefont{and}
  \bibinfo{author}{\bibfnamefont{O.~A.} \bibnamefont{Sampayo}},
  \bibinfo{journal}{Phys.Rev.} \textbf{\bibinfo{volume}{D84}},
  \bibinfo{pages}{113002} (\bibinfo{year}{2011}), \eprint{1110.0959}.

\bibitem[{\citenamefont{Blaksley et~al.}(2011)\citenamefont{Blaksley, Blennow,
  Bonnet, Coloma, and Fernandez-Martinez}}]{Blaksley:2011ey}
\bibinfo{author}{\bibfnamefont{C.}~\bibnamefont{Blaksley}},
  \bibinfo{author}{\bibfnamefont{M.}~\bibnamefont{Blennow}},
  \bibinfo{author}{\bibfnamefont{F.}~\bibnamefont{Bonnet}},
  \bibinfo{author}{\bibfnamefont{P.}~\bibnamefont{Coloma}}, \bibnamefont{and}
  \bibinfo{author}{\bibfnamefont{E.}~\bibnamefont{Fernandez-Martinez}},
  \bibinfo{journal}{Nucl.Phys.} \textbf{\bibinfo{volume}{B852}},
  \bibinfo{pages}{353} (\bibinfo{year}{2011}), \eprint{1105.0308}.

\bibitem[{\citenamefont{Duarte et~al.}(2015{\natexlab{a}})\citenamefont{Duarte,
  González-Sprinberg, and Sampayo}}]{Duarte:2014zea}
\bibinfo{author}{\bibfnamefont{L.}~\bibnamefont{Duarte}},
  \bibinfo{author}{\bibfnamefont{G.~A.} \bibnamefont{González-Sprinberg}},
  \bibnamefont{and} \bibinfo{author}{\bibfnamefont{O.~A.}
  \bibnamefont{Sampayo}}, \bibinfo{journal}{Phys. Rev.}
  \textbf{\bibinfo{volume}{D91}}, \bibinfo{pages}{053007}
  (\bibinfo{year}{2015}{\natexlab{a}}), \eprint{1412.1433}.

\bibitem[{\citenamefont{Antusch and Fischer}(2015)}]{Antusch:2015mia}
\bibinfo{author}{\bibfnamefont{S.}~\bibnamefont{Antusch}} \bibnamefont{and}
  \bibinfo{author}{\bibfnamefont{O.}~\bibnamefont{Fischer}},
  \bibinfo{journal}{JHEP} \textbf{\bibinfo{volume}{05}}, \bibinfo{pages}{053}
  (\bibinfo{year}{2015}), \eprint{1502.05915}.

\bibitem[{\citenamefont{Bray et~al.}(2005)\citenamefont{Bray, Lee, and
  Pilaftsis}}]{Bray:2005wv}
\bibinfo{author}{\bibfnamefont{S.}~\bibnamefont{Bray}},
  \bibinfo{author}{\bibfnamefont{J.~S.} \bibnamefont{Lee}}, \bibnamefont{and}
  \bibinfo{author}{\bibfnamefont{A.}~\bibnamefont{Pilaftsis}},
  \bibinfo{journal}{Phys.Lett.} \textbf{\bibinfo{volume}{B628}},
  \bibinfo{pages}{250} (\bibinfo{year}{2005}), \eprint{hep-ph/0508077}.

\bibitem[{\citenamefont{Peressutti et~al.}(2001)\citenamefont{Peressutti,
  Sampayo, and Aranda}}]{Peressutti:2001ms}
\bibinfo{author}{\bibfnamefont{J.}~\bibnamefont{Peressutti}},
  \bibinfo{author}{\bibfnamefont{O.}~\bibnamefont{Sampayo}}, \bibnamefont{and}
  \bibinfo{author}{\bibfnamefont{J.~I.} \bibnamefont{Aranda}},
  \bibinfo{journal}{Phys.Rev.} \textbf{\bibinfo{volume}{D64}},
  \bibinfo{pages}{073007} (\bibinfo{year}{2001}), \eprint{hep-ph/0105162}.

\bibitem[{\citenamefont{Peressutti and Sampayo}(2003)}]{Peressutti:2002nf}
\bibinfo{author}{\bibfnamefont{J.}~\bibnamefont{Peressutti}} \bibnamefont{and}
  \bibinfo{author}{\bibfnamefont{O.}~\bibnamefont{Sampayo}},
  \bibinfo{journal}{Phys.Rev.} \textbf{\bibinfo{volume}{D67}},
  \bibinfo{pages}{017302} (\bibinfo{year}{2003}), \eprint{hep-ph/0211355}.

\bibitem[{\citenamefont{Antusch
  et~al.}(2017{\natexlab{a}})\citenamefont{Antusch, Cazzato, and
  Fischer}}]{Antusch:2016ejd}
\bibinfo{author}{\bibfnamefont{S.}~\bibnamefont{Antusch}},
  \bibinfo{author}{\bibfnamefont{E.}~\bibnamefont{Cazzato}}, \bibnamefont{and}
  \bibinfo{author}{\bibfnamefont{O.}~\bibnamefont{Fischer}},
  \bibinfo{journal}{Int. J. Mod. Phys.} \textbf{\bibinfo{volume}{A32}},
  \bibinfo{pages}{1750078} (\bibinfo{year}{2017}{\natexlab{a}}),
  \eprint{1612.02728}.

\bibitem[{\citenamefont{Khachatryan
  et~al.}(2016{\natexlab{a}})}]{Khachatryan:2016olu}
\bibinfo{author}{\bibfnamefont{V.}~\bibnamefont{Khachatryan}}
  \bibnamefont{et~al.} (\bibinfo{collaboration}{CMS}), \bibinfo{journal}{JHEP}
  \textbf{\bibinfo{volume}{04}}, \bibinfo{pages}{169}
  (\bibinfo{year}{2016}{\natexlab{a}}), \eprint{1603.02248}.

\bibitem[{\citenamefont{Khachatryan
  et~al.}(2015{\natexlab{a}})}]{Khachatryan:2015gha}
\bibinfo{author}{\bibfnamefont{V.}~\bibnamefont{Khachatryan}}
  \bibnamefont{et~al.} (\bibinfo{collaboration}{CMS}), \bibinfo{journal}{Phys.
  Lett.} \textbf{\bibinfo{volume}{B748}}, \bibinfo{pages}{144}
  (\bibinfo{year}{2015}{\natexlab{a}}), \eprint{1501.05566}.

\bibitem[{\citenamefont{Aad et~al.}(2015{\natexlab{a}})}]{Aad:2015xaa}
\bibinfo{author}{\bibfnamefont{G.}~\bibnamefont{Aad}} \bibnamefont{et~al.}
  (\bibinfo{collaboration}{ATLAS}), \bibinfo{journal}{JHEP}
  \textbf{\bibinfo{volume}{07}}, \bibinfo{pages}{162}
  (\bibinfo{year}{2015}{\natexlab{a}}), \eprint{1506.06020}.

\bibitem[{\citenamefont{Aad et~al.}(2011)}]{Aad:2011vj}
\bibinfo{author}{\bibfnamefont{G.}~\bibnamefont{Aad}} \bibnamefont{et~al.}
  (\bibinfo{collaboration}{ATLAS Collaboration}), \bibinfo{journal}{JHEP}
  \textbf{\bibinfo{volume}{1110}}, \bibinfo{pages}{107} (\bibinfo{year}{2011}),
  \eprint{1108.0366}.

\bibitem[{\citenamefont{Aad et~al.}(2012)}]{ATLAS:2012ak}
\bibinfo{author}{\bibfnamefont{G.}~\bibnamefont{Aad}} \bibnamefont{et~al.}
  (\bibinfo{collaboration}{ATLAS Collaboration}),
  \bibinfo{journal}{Eur.Phys.J.} \textbf{\bibinfo{volume}{C72}},
  \bibinfo{pages}{2056} (\bibinfo{year}{2012}), \eprint{1203.5420}.

\bibitem[{\citenamefont{Aparici et~al.}(2009)\citenamefont{Aparici, Kim,
  Santamaria, and Wudka}}]{Aparici:2009fh}
\bibinfo{author}{\bibfnamefont{A.}~\bibnamefont{Aparici}},
  \bibinfo{author}{\bibfnamefont{K.}~\bibnamefont{Kim}},
  \bibinfo{author}{\bibfnamefont{A.}~\bibnamefont{Santamaria}},
  \bibnamefont{and} \bibinfo{author}{\bibfnamefont{J.}~\bibnamefont{Wudka}},
  \bibinfo{journal}{Phys. Rev.} \textbf{\bibinfo{volume}{D80}},
  \bibinfo{pages}{013010} (\bibinfo{year}{2009}), \eprint{0904.3244}.

\bibitem[{\citenamefont{Ballett et~al.}(2017)\citenamefont{Ballett, Pascoli,
  and Ross-Lonergan}}]{Ballett:2016opr}
\bibinfo{author}{\bibfnamefont{P.}~\bibnamefont{Ballett}},
  \bibinfo{author}{\bibfnamefont{S.}~\bibnamefont{Pascoli}}, \bibnamefont{and}
  \bibinfo{author}{\bibfnamefont{M.}~\bibnamefont{Ross-Lonergan}},
  \bibinfo{journal}{JHEP} \textbf{\bibinfo{volume}{04}}, \bibinfo{pages}{102}
  (\bibinfo{year}{2017}), \eprint{1610.08512}.

\bibitem[{\citenamefont{Caputo et~al.}(2017)\citenamefont{Caputo, Hernandez,
  Lopez-Pavon, and Salvado}}]{Caputo:2017pit}
\bibinfo{author}{\bibfnamefont{A.}~\bibnamefont{Caputo}},
  \bibinfo{author}{\bibfnamefont{P.}~\bibnamefont{Hernandez}},
  \bibinfo{author}{\bibfnamefont{J.}~\bibnamefont{Lopez-Pavon}},
  \bibnamefont{and} \bibinfo{author}{\bibfnamefont{J.}~\bibnamefont{Salvado}},
  \bibinfo{journal}{JHEP} \textbf{\bibinfo{volume}{06}}, \bibinfo{pages}{112}
  (\bibinfo{year}{2017}), \eprint{1704.08721}.

\bibitem[{\citenamefont{Bhattacharya and Wudka}(2016)}]{Bhattacharya:2015vja}
\bibinfo{author}{\bibfnamefont{S.}~\bibnamefont{Bhattacharya}}
  \bibnamefont{and} \bibinfo{author}{\bibfnamefont{J.}~\bibnamefont{Wudka}},
  \bibinfo{journal}{Phys. Rev.} \textbf{\bibinfo{volume}{D94}},
  \bibinfo{pages}{055022} (\bibinfo{year}{2016}), \bibinfo{note}{[Erratum:
  Phys. Rev.D95,no.3,039904(2017)]}, \eprint{1505.05264}.

\bibitem[{\citenamefont{Liao and Ma}(2017)}]{Liao:2016qyd}
\bibinfo{author}{\bibfnamefont{Y.}~\bibnamefont{Liao}} \bibnamefont{and}
  \bibinfo{author}{\bibfnamefont{X.-D.} \bibnamefont{Ma}},
  \bibinfo{journal}{Phys. Rev.} \textbf{\bibinfo{volume}{D96}},
  \bibinfo{pages}{015012} (\bibinfo{year}{2017}), \eprint{1612.04527}.

\bibitem[{\citenamefont{Yue et~al.}(2017)\citenamefont{Yue, Guo, and
  Zhao}}]{Yue:2017mmi}
\bibinfo{author}{\bibfnamefont{C.-X.} \bibnamefont{Yue}},
  \bibinfo{author}{\bibfnamefont{Y.-C.} \bibnamefont{Guo}}, \bibnamefont{and}
  \bibinfo{author}{\bibfnamefont{Z.-H.} \bibnamefont{Zhao}}
  (\bibinfo{year}{2017}), \eprint{1710.06144}.

\bibitem[{\citenamefont{Dib et~al.}(2016)\citenamefont{Dib, Kim, Wang, and
  Zhang}}]{Dib:2016wge}
\bibinfo{author}{\bibfnamefont{C.~O.} \bibnamefont{Dib}},
  \bibinfo{author}{\bibfnamefont{C.~S.} \bibnamefont{Kim}},
  \bibinfo{author}{\bibfnamefont{K.}~\bibnamefont{Wang}}, \bibnamefont{and}
  \bibinfo{author}{\bibfnamefont{J.}~\bibnamefont{Zhang}},
  \bibinfo{journal}{Phys. Rev.} \textbf{\bibinfo{volume}{D94}},
  \bibinfo{pages}{013005} (\bibinfo{year}{2016}), \eprint{1605.01123}.

\bibitem[{\citenamefont{Batell et~al.}(2016)\citenamefont{Batell, Pospelov, and
  Shuve}}]{Batell:2016zod}
\bibinfo{author}{\bibfnamefont{B.}~\bibnamefont{Batell}},
  \bibinfo{author}{\bibfnamefont{M.}~\bibnamefont{Pospelov}}, \bibnamefont{and}
  \bibinfo{author}{\bibfnamefont{B.}~\bibnamefont{Shuve}},
  \bibinfo{journal}{JHEP} \textbf{\bibinfo{volume}{08}}, \bibinfo{pages}{052}
  (\bibinfo{year}{2016}), \eprint{1604.06099}.

\bibitem[{\citenamefont{Helo et~al.}(2014)\citenamefont{Helo, Hirsch, and
  Kovalenko}}]{Helo:2013esa}
\bibinfo{author}{\bibfnamefont{J.~C.} \bibnamefont{Helo}},
  \bibinfo{author}{\bibfnamefont{M.}~\bibnamefont{Hirsch}}, \bibnamefont{and}
  \bibinfo{author}{\bibfnamefont{S.}~\bibnamefont{Kovalenko}},
  \bibinfo{journal}{Phys. Rev.} \textbf{\bibinfo{volume}{D89}},
  \bibinfo{pages}{073005} (\bibinfo{year}{2014}), \eprint{1312.2900}.

\bibitem[{\citenamefont{Gago et~al.}(2015)\citenamefont{Gago, Hernández,
  Jones-Pérez, Losada, and Briceño}}]{Gago:2015vma}
\bibinfo{author}{\bibfnamefont{A.~M.} \bibnamefont{Gago}},
  \bibinfo{author}{\bibfnamefont{P.}~\bibnamefont{Hernández}},
  \bibinfo{author}{\bibfnamefont{J.}~\bibnamefont{Jones-Pérez}},
  \bibinfo{author}{\bibfnamefont{M.}~\bibnamefont{Losada}}, \bibnamefont{and}
  \bibinfo{author}{\bibfnamefont{A.~M.} \bibnamefont{Briceño}},
  \bibinfo{journal}{Eur. Phys. J.} \textbf{\bibinfo{volume}{C75}},
  \bibinfo{pages}{470} (\bibinfo{year}{2015}), \eprint{1505.05880}.

\bibitem[{\citenamefont{Cerdeño et~al.}(2014)\citenamefont{Cerdeño,
  Martín-Lozano, and Seto}}]{Cerdeno:2013oya}
\bibinfo{author}{\bibfnamefont{D.~G.} \bibnamefont{Cerdeño}},
  \bibinfo{author}{\bibfnamefont{V.}~\bibnamefont{Martín-Lozano}},
  \bibnamefont{and} \bibinfo{author}{\bibfnamefont{O.}~\bibnamefont{Seto}},
  \bibinfo{journal}{JHEP} \textbf{\bibinfo{volume}{05}}, \bibinfo{pages}{035}
  (\bibinfo{year}{2014}), \eprint{1311.7260}.

\bibitem[{\citenamefont{Antusch
  et~al.}(2017{\natexlab{b}})\citenamefont{Antusch, Cazzato, and
  Fischer}}]{Antusch:2017hhu}
\bibinfo{author}{\bibfnamefont{S.}~\bibnamefont{Antusch}},
  \bibinfo{author}{\bibfnamefont{E.}~\bibnamefont{Cazzato}}, \bibnamefont{and}
  \bibinfo{author}{\bibfnamefont{O.}~\bibnamefont{Fischer}},
  \bibinfo{journal}{Phys. Lett.} \textbf{\bibinfo{volume}{B774}},
  \bibinfo{pages}{114} (\bibinfo{year}{2017}{\natexlab{b}}),
  \eprint{1706.05990}.

\bibitem[{\citenamefont{Blondel et~al.}(2016)\citenamefont{Blondel, Graverini,
  Serra, and Shaposhnikov}}]{Blondel:2014bra}
\bibinfo{author}{\bibfnamefont{A.}~\bibnamefont{Blondel}},
  \bibinfo{author}{\bibfnamefont{E.}~\bibnamefont{Graverini}},
  \bibinfo{author}{\bibfnamefont{N.}~\bibnamefont{Serra}}, \bibnamefont{and}
  \bibinfo{author}{\bibfnamefont{M.}~\bibnamefont{Shaposhnikov}}
  (\bibinfo{collaboration}{FCC-ee study Team}), in
  \emph{\bibinfo{booktitle}{{Proceedings, 37th International Conference on High
  Energy Physics (ICHEP 2014): Valencia, Spain, July 2-9, 2014}}}
  (\bibinfo{year}{2016}), \eprint{1411.5230},
  \urlprefix\url{http://inspirehep.net/record/1328783/files/arXiv:1411.5230.pdf}.

\bibitem[{\citenamefont{Antusch et~al.}(2016)\citenamefont{Antusch, Cazzato,
  and Fischer}}]{Antusch:2016vyf}
\bibinfo{author}{\bibfnamefont{S.}~\bibnamefont{Antusch}},
  \bibinfo{author}{\bibfnamefont{E.}~\bibnamefont{Cazzato}}, \bibnamefont{and}
  \bibinfo{author}{\bibfnamefont{O.}~\bibnamefont{Fischer}},
  \bibinfo{journal}{JHEP} \textbf{\bibinfo{volume}{12}}, \bibinfo{pages}{007}
  (\bibinfo{year}{2016}), \eprint{1604.02420}.

\bibitem[{\citenamefont{Duarte et~al.}(2015{\natexlab{b}})\citenamefont{Duarte,
  Peressutti, and Sampayo}}]{Duarte:2015iba}
\bibinfo{author}{\bibfnamefont{L.}~\bibnamefont{Duarte}},
  \bibinfo{author}{\bibfnamefont{J.}~\bibnamefont{Peressutti}},
  \bibnamefont{and} \bibinfo{author}{\bibfnamefont{O.~A.}
  \bibnamefont{Sampayo}}, \bibinfo{journal}{Phys. Rev.}
  \textbf{\bibinfo{volume}{D92}}, \bibinfo{pages}{093002}
  (\bibinfo{year}{2015}{\natexlab{b}}), \eprint{1508.01588}.

\bibitem[{\citenamefont{Aguilar-Arevalo et~al.}(2007)}]{AguilarArevalo:2007it}
\bibinfo{author}{\bibfnamefont{A.~A.} \bibnamefont{Aguilar-Arevalo}}
  \bibnamefont{et~al.} (\bibinfo{collaboration}{MiniBooNE}),
  \bibinfo{journal}{Phys. Rev. Lett.} \textbf{\bibinfo{volume}{98}},
  \bibinfo{pages}{231801} (\bibinfo{year}{2007}), \eprint{0704.1500}.

\bibitem[{\citenamefont{Aguilar-Arevalo et~al.}(2009)}]{AguilarArevalo:2008rc}
\bibinfo{author}{\bibfnamefont{A.~A.} \bibnamefont{Aguilar-Arevalo}}
  \bibnamefont{et~al.} (\bibinfo{collaboration}{MiniBooNE}),
  \bibinfo{journal}{Phys. Rev. Lett.} \textbf{\bibinfo{volume}{102}},
  \bibinfo{pages}{101802} (\bibinfo{year}{2009}), \eprint{0812.2243}.

\bibitem[{\citenamefont{Sinitsyna et~al.}(2013)\citenamefont{Sinitsyna, Masip,
  and Sinitsyna}}]{Sinitsyna:2013hmn}
\bibinfo{author}{\bibfnamefont{V.~G.} \bibnamefont{Sinitsyna}},
  \bibinfo{author}{\bibfnamefont{M.}~\bibnamefont{Masip}}, \bibnamefont{and}
  \bibinfo{author}{\bibfnamefont{V.~Y.} \bibnamefont{Sinitsyna}},
  \bibinfo{journal}{EPJ Web Conf.} \textbf{\bibinfo{volume}{52}},
  \bibinfo{pages}{09010} (\bibinfo{year}{2013}).

\bibitem[{\citenamefont{Gninenko}(2009)}]{Gninenko:2009ks}
\bibinfo{author}{\bibfnamefont{S.~N.} \bibnamefont{Gninenko}},
  \bibinfo{journal}{Phys. Rev. Lett.} \textbf{\bibinfo{volume}{103}},
  \bibinfo{pages}{241802} (\bibinfo{year}{2009}), \eprint{0902.3802}.

\bibitem[{\citenamefont{Aad et~al.}(2014)}]{Aad:2014gfa}
\bibinfo{author}{\bibfnamefont{G.}~\bibnamefont{Aad}} \bibnamefont{et~al.}
  (\bibinfo{collaboration}{ATLAS}), \bibinfo{journal}{Phys. Rev.}
  \textbf{\bibinfo{volume}{D90}}, \bibinfo{pages}{112005}
  (\bibinfo{year}{2014}), \eprint{1409.5542}.

\bibitem[{\citenamefont{Aad et~al.}(2013)}]{Aad:2013oua}
\bibinfo{author}{\bibfnamefont{G.}~\bibnamefont{Aad}} \bibnamefont{et~al.}
  (\bibinfo{collaboration}{ATLAS}), \bibinfo{journal}{Phys. Rev.}
  \textbf{\bibinfo{volume}{D88}}, \bibinfo{pages}{012001}
  (\bibinfo{year}{2013}), \eprint{1304.6310}.

\bibitem[{\citenamefont{Khachatryan et~al.}(2016{\natexlab{b}})}]{CMS:2015loa}
\bibinfo{author}{\bibfnamefont{V.}~\bibnamefont{Khachatryan}}
  \bibnamefont{et~al.} (\bibinfo{collaboration}{CMS}), \bibinfo{journal}{Phys.
  Lett.} \textbf{\bibinfo{volume}{B757}}, \bibinfo{pages}{6}
  (\bibinfo{year}{2016}{\natexlab{b}}), \eprint{1508.01218}.

\bibitem[{\citenamefont{Wudka}(2000)}]{Wudka:1999ax}
\bibinfo{author}{\bibfnamefont{J.}~\bibnamefont{Wudka}}, \bibinfo{journal}{AIP
  Conf.Proc.} \textbf{\bibinfo{volume}{531}}, \bibinfo{pages}{81}
  (\bibinfo{year}{2000}), \eprint{hep-ph/0002180}.

\bibitem[{\citenamefont{Weinberg}(1979)}]{Weinberg:1979sa}
\bibinfo{author}{\bibfnamefont{S.}~\bibnamefont{Weinberg}},
  \bibinfo{journal}{Phys. Rev. Lett.} \textbf{\bibinfo{volume}{43}},
  \bibinfo{pages}{1566} (\bibinfo{year}{1979}).

\bibitem[{\citenamefont{Arzt et~al.}(1995)\citenamefont{Arzt, Einhorn, and
  Wudka}}]{Arzt:1994gp}
\bibinfo{author}{\bibfnamefont{C.}~\bibnamefont{Arzt}},
  \bibinfo{author}{\bibfnamefont{M.}~\bibnamefont{Einhorn}}, \bibnamefont{and}
  \bibinfo{author}{\bibfnamefont{J.}~\bibnamefont{Wudka}},
  \bibinfo{journal}{Nucl.Phys.} \textbf{\bibinfo{volume}{B433}},
  \bibinfo{pages}{41} (\bibinfo{year}{1995}), \eprint{hep-ph/9405214}.

\bibitem[{\citenamefont{Duarte et~al.}(2016)\citenamefont{Duarte, Romero,
  Peressutti, and Sampayo}}]{Duarte:2016miz}
\bibinfo{author}{\bibfnamefont{L.}~\bibnamefont{Duarte}},
  \bibinfo{author}{\bibfnamefont{I.}~\bibnamefont{Romero}},
  \bibinfo{author}{\bibfnamefont{J.}~\bibnamefont{Peressutti}},
  \bibnamefont{and} \bibinfo{author}{\bibfnamefont{O.~A.}
  \bibnamefont{Sampayo}}, \bibinfo{journal}{Eur. Phys. J.}
  \textbf{\bibinfo{volume}{C76}}, \bibinfo{pages}{453} (\bibinfo{year}{2016}),
  \eprint{1603.08052}.

\bibitem[{\citenamefont{Ruiz}(2017)}]{Ruiz:2017nip}
\bibinfo{author}{\bibfnamefont{R.}~\bibnamefont{Ruiz}}, \bibinfo{journal}{Eur.
  Phys. J.} \textbf{\bibinfo{volume}{C77}}, \bibinfo{pages}{375}
  (\bibinfo{year}{2017}), \eprint{1703.04669}.

\bibitem[{\citenamefont{de~Gouvêa and Kobach}(2016)}]{deGouvea:2015euy}
\bibinfo{author}{\bibfnamefont{A.}~\bibnamefont{de~Gouvêa}} \bibnamefont{and}
  \bibinfo{author}{\bibfnamefont{A.}~\bibnamefont{Kobach}},
  \bibinfo{journal}{Phys. Rev.} \textbf{\bibinfo{volume}{D93}},
  \bibinfo{pages}{033005} (\bibinfo{year}{2016}), \eprint{1511.00683}.

\bibitem[{\citenamefont{Deppisch et~al.}(2015)\citenamefont{Deppisch,
  Bhupal~Dev, and Pilaftsis}}]{Deppisch:2015qwa}
\bibinfo{author}{\bibfnamefont{F.~F.} \bibnamefont{Deppisch}},
  \bibinfo{author}{\bibfnamefont{P.~S.} \bibnamefont{Bhupal~Dev}},
  \bibnamefont{and}
  \bibinfo{author}{\bibfnamefont{A.}~\bibnamefont{Pilaftsis}},
  \bibinfo{journal}{New J. Phys.} \textbf{\bibinfo{volume}{17}},
  \bibinfo{pages}{075019} (\bibinfo{year}{2015}), \eprint{1502.06541}.

\bibitem[{\citenamefont{Das et~al.}(2014)\citenamefont{Das, Bhupal~Dev, and
  Okada}}]{Das:2014jxa}
\bibinfo{author}{\bibfnamefont{A.}~\bibnamefont{Das}},
  \bibinfo{author}{\bibfnamefont{P.~S.} \bibnamefont{Bhupal~Dev}},
  \bibnamefont{and} \bibinfo{author}{\bibfnamefont{N.}~\bibnamefont{Okada}},
  \bibinfo{journal}{Phys. Lett.} \textbf{\bibinfo{volume}{B735}},
  \bibinfo{pages}{364} (\bibinfo{year}{2014}), \eprint{1405.0177}.

\bibitem[{\citenamefont{Fernandez-Martinez
  et~al.}(2016)\citenamefont{Fernandez-Martinez, Hernandez-Garcia, and
  Lopez-Pavon}}]{Fernandez-Martinez:2016lgt}
\bibinfo{author}{\bibfnamefont{E.}~\bibnamefont{Fernandez-Martinez}},
  \bibinfo{author}{\bibfnamefont{J.}~\bibnamefont{Hernandez-Garcia}},
  \bibnamefont{and}
  \bibinfo{author}{\bibfnamefont{J.}~\bibnamefont{Lopez-Pavon}},
  \bibinfo{journal}{JHEP} \textbf{\bibinfo{volume}{08}}, \bibinfo{pages}{033}
  (\bibinfo{year}{2016}), \eprint{1605.08774}.

\bibitem[{\citenamefont{Gando et~al.}(2016)}]{KamLAND-Zen:2016pfg}
\bibinfo{author}{\bibfnamefont{A.}~\bibnamefont{Gando}} \bibnamefont{et~al.}
  (\bibinfo{collaboration}{KamLAND-Zen}), \bibinfo{journal}{Phys. Rev. Lett.}
  \textbf{\bibinfo{volume}{117}}, \bibinfo{pages}{082503}
  (\bibinfo{year}{2016}), \bibinfo{note}{[Addendum: Phys. Rev.
  Lett.117,no.10,109903(2016)]}, \eprint{1605.02889}.

\bibitem[{\citenamefont{Mohapatra}(1999)}]{Mohapatra:1998ye}
\bibinfo{author}{\bibfnamefont{R.}~\bibnamefont{Mohapatra}},
  \bibinfo{journal}{Nucl.Phys.Proc.Suppl.} \textbf{\bibinfo{volume}{77}},
  \bibinfo{pages}{376} (\bibinfo{year}{1999}), \eprint{hep-ph/9808284}.

\bibitem[{\citenamefont{Abreu et~al.}(1997)}]{Abreu:1996pa}
\bibinfo{author}{\bibfnamefont{P.}~\bibnamefont{Abreu}} \bibnamefont{et~al.}
  (\bibinfo{collaboration}{DELPHI}), \bibinfo{journal}{Z. Phys.}
  \textbf{\bibinfo{volume}{C74}}, \bibinfo{pages}{57} (\bibinfo{year}{1997}),
  \bibinfo{note}{[Erratum: Z. Phys.C75,580(1997)]}.

\bibitem[{\citenamefont{Liventsev et~al.}(2013)}]{Liventsev:2013zz}
\bibinfo{author}{\bibfnamefont{D.}~\bibnamefont{Liventsev}}
  \bibnamefont{et~al.} (\bibinfo{collaboration}{Belle}),
  \bibinfo{journal}{Phys. Rev.} \textbf{\bibinfo{volume}{D87}},
  \bibinfo{pages}{071102} (\bibinfo{year}{2013}), \eprint{1301.1105}.

\bibitem[{\citenamefont{Aaij et~al.}(2014)}]{Aaij:2014aba}
\bibinfo{author}{\bibfnamefont{R.}~\bibnamefont{Aaij}} \bibnamefont{et~al.}
  (\bibinfo{collaboration}{LHCb}), \bibinfo{journal}{Phys. Rev. Lett.}
  \textbf{\bibinfo{volume}{112}}, \bibinfo{pages}{131802}
  (\bibinfo{year}{2014}), \eprint{1401.5361}.

\bibitem[{\citenamefont{Shuve and Peskin}(2016)}]{Shuve:2016muy}
\bibinfo{author}{\bibfnamefont{B.}~\bibnamefont{Shuve}} \bibnamefont{and}
  \bibinfo{author}{\bibfnamefont{M.~E.} \bibnamefont{Peskin}},
  \bibinfo{journal}{Phys. Rev.} \textbf{\bibinfo{volume}{D94}},
  \bibinfo{pages}{113007} (\bibinfo{year}{2016}), \eprint{1607.04258}.

\bibitem[{\citenamefont{Cvetic and Kim}(2016)}]{Cvetic:2016fbv}
\bibinfo{author}{\bibfnamefont{G.}~\bibnamefont{Cvetic}} \bibnamefont{and}
  \bibinfo{author}{\bibfnamefont{C.~S.} \bibnamefont{Kim}},
  \bibinfo{journal}{Phys. Rev.} \textbf{\bibinfo{volume}{D94}},
  \bibinfo{pages}{053001} (\bibinfo{year}{2016}), \eprint{1606.04140}.

\bibitem[{\citenamefont{Milanes et~al.}(2016)\citenamefont{Milanes, Quintero,
  and Vera}}]{Milanes:2016rzr}
\bibinfo{author}{\bibfnamefont{D.}~\bibnamefont{Milanes}},
  \bibinfo{author}{\bibfnamefont{N.}~\bibnamefont{Quintero}}, \bibnamefont{and}
  \bibinfo{author}{\bibfnamefont{C.~E.} \bibnamefont{Vera}},
  \bibinfo{journal}{Phys. Rev.} \textbf{\bibinfo{volume}{D93}},
  \bibinfo{pages}{094026} (\bibinfo{year}{2016}), \eprint{1604.03177}.

\bibitem[{\citenamefont{Mandal and Sinha}(2016)}]{Mandal:2016hpr}
\bibinfo{author}{\bibfnamefont{S.}~\bibnamefont{Mandal}} \bibnamefont{and}
  \bibinfo{author}{\bibfnamefont{N.}~\bibnamefont{Sinha}},
  \bibinfo{journal}{Phys. Rev.} \textbf{\bibinfo{volume}{D94}},
  \bibinfo{pages}{033001} (\bibinfo{year}{2016}), \eprint{1602.09112}.

\bibitem[{\citenamefont{Dib and Kim}(2014)}]{Dib:2014iga}
\bibinfo{author}{\bibfnamefont{C.}~\bibnamefont{Dib}} \bibnamefont{and}
  \bibinfo{author}{\bibfnamefont{C.~S.} \bibnamefont{Kim}},
  \bibinfo{journal}{Phys. Rev.} \textbf{\bibinfo{volume}{D89}},
  \bibinfo{pages}{077301} (\bibinfo{year}{2014}), \eprint{1403.1985}.

\bibitem[{\citenamefont{Tommasini et~al.}(1995)\citenamefont{Tommasini,
  Barenboim, Bernabeu, and Jarlskog}}]{Tommasini:1995ii}
\bibinfo{author}{\bibfnamefont{D.}~\bibnamefont{Tommasini}},
  \bibinfo{author}{\bibfnamefont{G.}~\bibnamefont{Barenboim}},
  \bibinfo{author}{\bibfnamefont{J.}~\bibnamefont{Bernabeu}}, \bibnamefont{and}
  \bibinfo{author}{\bibfnamefont{C.}~\bibnamefont{Jarlskog}},
  \bibinfo{journal}{Nucl.Phys.} \textbf{\bibinfo{volume}{B444}},
  \bibinfo{pages}{451} (\bibinfo{year}{1995}), \eprint{hep-ph/9503228}.

\bibitem[{\citenamefont{del Aguila et~al.}(2008)\citenamefont{del Aguila,
  de~Blas, and Perez-Victoria}}]{delAguila:2008pw}
\bibinfo{author}{\bibfnamefont{F.}~\bibnamefont{del Aguila}},
  \bibinfo{author}{\bibfnamefont{J.}~\bibnamefont{de~Blas}}, \bibnamefont{and}
  \bibinfo{author}{\bibfnamefont{M.}~\bibnamefont{Perez-Victoria}},
  \bibinfo{journal}{Phys. Rev.} \textbf{\bibinfo{volume}{D78}},
  \bibinfo{pages}{013010} (\bibinfo{year}{2008}), \eprint{0803.4008}.

\bibitem[{\citenamefont{Pumplin et~al.}(2002)\citenamefont{Pumplin, Stump,
  Huston, Lai, Nadolsky, and Tung}}]{Pumplin:2002vw}
\bibinfo{author}{\bibfnamefont{J.}~\bibnamefont{Pumplin}},
  \bibinfo{author}{\bibfnamefont{D.~R.} \bibnamefont{Stump}},
  \bibinfo{author}{\bibfnamefont{J.}~\bibnamefont{Huston}},
  \bibinfo{author}{\bibfnamefont{H.~L.} \bibnamefont{Lai}},
  \bibinfo{author}{\bibfnamefont{P.~M.} \bibnamefont{Nadolsky}},
  \bibnamefont{and} \bibinfo{author}{\bibfnamefont{W.~K.} \bibnamefont{Tung}},
  \bibinfo{journal}{JHEP} \textbf{\bibinfo{volume}{07}}, \bibinfo{pages}{012}
  (\bibinfo{year}{2002}), \eprint{hep-ph/0201195}.

\bibitem[{\citenamefont{Khachatryan
  et~al.}(2015{\natexlab{b}})}]{Khachatryan:2014mea}
\bibinfo{author}{\bibfnamefont{V.}~\bibnamefont{Khachatryan}}
  \bibnamefont{et~al.} (\bibinfo{collaboration}{CMS}), \bibinfo{journal}{Phys.
  Rev. Lett.} \textbf{\bibinfo{volume}{114}}, \bibinfo{pages}{061801}
  (\bibinfo{year}{2015}{\natexlab{b}}), \eprint{1409.4789}.

\bibitem[{\citenamefont{Khachatryan et~al.}(2015{\natexlab{c}})}]{CMS:2014hka}
\bibinfo{author}{\bibfnamefont{V.}~\bibnamefont{Khachatryan}}
  \bibnamefont{et~al.} (\bibinfo{collaboration}{CMS}), \bibinfo{journal}{Phys.
  Rev.} \textbf{\bibinfo{volume}{D91}}, \bibinfo{pages}{052012}
  (\bibinfo{year}{2015}{\natexlab{c}}), \eprint{1411.6977}.

\bibitem[{\citenamefont{Aad et~al.}(2015{\natexlab{b}})}]{Aad:2015rba}
\bibinfo{author}{\bibfnamefont{G.}~\bibnamefont{Aad}} \bibnamefont{et~al.}
  (\bibinfo{collaboration}{ATLAS}), \bibinfo{journal}{Phys. Rev.}
  \textbf{\bibinfo{volume}{D92}}, \bibinfo{pages}{072004}
  (\bibinfo{year}{2015}{\natexlab{b}}), \eprint{1504.05162}.

\bibitem[{\citenamefont{Aad et~al.}(2015{\natexlab{c}})}]{Aad:2015uaa}
\bibinfo{author}{\bibfnamefont{G.}~\bibnamefont{Aad}} \bibnamefont{et~al.}
  (\bibinfo{collaboration}{ATLAS}), \bibinfo{journal}{Phys. Rev.}
  \textbf{\bibinfo{volume}{D92}}, \bibinfo{pages}{012010}
  (\bibinfo{year}{2015}{\natexlab{c}}), \eprint{1504.03634}.

\bibitem[{\citenamefont{Evans and Shelton}(2016)}]{Evans:2016zau}
\bibinfo{author}{\bibfnamefont{J.~A.} \bibnamefont{Evans}} \bibnamefont{and}
  \bibinfo{author}{\bibfnamefont{J.}~\bibnamefont{Shelton}},
  \bibinfo{journal}{JHEP} \textbf{\bibinfo{volume}{04}}, \bibinfo{pages}{056}
  (\bibinfo{year}{2016}), \eprint{1601.01326}.

\bibitem[{\citenamefont{Kleiss et~al.}(1986)\citenamefont{Kleiss, Stirling, and
  Ellis}}]{Kleiss:1985gy}
\bibinfo{author}{\bibfnamefont{R.}~\bibnamefont{Kleiss}},
  \bibinfo{author}{\bibfnamefont{W.~J.} \bibnamefont{Stirling}},
  \bibnamefont{and} \bibinfo{author}{\bibfnamefont{S.}~\bibnamefont{Ellis}},
  \bibinfo{journal}{Comput.Phys.Commun.} \textbf{\bibinfo{volume}{40}},
  \bibinfo{pages}{359} (\bibinfo{year}{1986}).

\bibitem[{\citenamefont{Uhlemann and Kauer}(2009)}]{Uhlemann:2008pm}
\bibinfo{author}{\bibfnamefont{C.~F.} \bibnamefont{Uhlemann}} \bibnamefont{and}
  \bibinfo{author}{\bibfnamefont{N.}~\bibnamefont{Kauer}},
  \bibinfo{journal}{Nucl. Phys.} \textbf{\bibinfo{volume}{B814}},
  \bibinfo{pages}{195} (\bibinfo{year}{2009}), \eprint{0807.4112}.

\bibitem[{\citenamefont{Khachatryan
  et~al.}(2016{\natexlab{c}})}]{Khachatryan:2016hns}
\bibinfo{author}{\bibfnamefont{V.}~\bibnamefont{Khachatryan}}
  \bibnamefont{et~al.} (\bibinfo{collaboration}{CMS}), \bibinfo{journal}{Phys.
  Lett.} \textbf{\bibinfo{volume}{B759}}, \bibinfo{pages}{479}
  (\bibinfo{year}{2016}{\natexlab{c}}), \eprint{1602.08772}.

\bibitem[{\citenamefont{Belyaev et~al.}(2013)\citenamefont{Belyaev,
  Christensen, and Pukhov}}]{Belyaev:2012qa}
\bibinfo{author}{\bibfnamefont{A.}~\bibnamefont{Belyaev}},
  \bibinfo{author}{\bibfnamefont{N.~D.} \bibnamefont{Christensen}},
  \bibnamefont{and} \bibinfo{author}{\bibfnamefont{A.}~\bibnamefont{Pukhov}},
  \bibinfo{journal}{Comput.Phys.Commun.} \textbf{\bibinfo{volume}{184}},
  \bibinfo{pages}{1729} (\bibinfo{year}{2013}), \eprint{1207.6082}.

\bibitem[{ATL(2015)}]{ATL-PHYS-PUB-2015-027}
\bibinfo{type}{Tech. Rep.} \bibinfo{number}{ATL-PHYS-PUB-2015-027},
  \bibinfo{institution}{CERN}, \bibinfo{address}{Geneva}
  (\bibinfo{year}{2015}), \urlprefix\url{http://cds.cern.ch/record/2037904}.

\bibitem[{\citenamefont{Aad et~al.}(2017)}]{Aad:2016nrq}
\bibinfo{author}{\bibfnamefont{G.}~\bibnamefont{Aad}} \bibnamefont{et~al.}
  (\bibinfo{collaboration}{ATLAS}), \bibinfo{journal}{Eur. Phys. J.}
  \textbf{\bibinfo{volume}{C77}}, \bibinfo{pages}{241} (\bibinfo{year}{2017}),
  \eprint{1609.09324}.

\bibitem[{\citenamefont{Khachatryan
  et~al.}(2015{\natexlab{d}})}]{Khachatryan:2014gga}
\bibinfo{author}{\bibfnamefont{V.}~\bibnamefont{Khachatryan}}
  \bibnamefont{et~al.} (\bibinfo{collaboration}{CMS}), \bibinfo{journal}{JINST}
  \textbf{\bibinfo{volume}{10}}, \bibinfo{pages}{P02006}
  (\bibinfo{year}{2015}{\natexlab{d}}), \eprint{1411.0511}.

\bibitem[{\citenamefont{Nikiforou}(2014-02-26)}]{Nikiforou:2014cka}
\bibinfo{author}{\bibfnamefont{N.}~\bibnamefont{Nikiforou}}, Ph.D. thesis,
  \bibinfo{school}{Columbia U.} (\bibinfo{year}{2014-02-26}),
  \urlprefix\url{http://inspirehep.net/record/1296368/files/594721125_CERN-THESIS-2014-011.pdf}.

\end{thebibliography}

\end{document}